\documentclass{article}

\usepackage{PRIMEarxiv}

\usepackage[utf8]{inputenc} 
\usepackage[T1]{fontenc}    
\usepackage{url}            
\usepackage{booktabs,longtable,array,ragged2e,xcolor,colortbl}
\usepackage{amsfonts}       
\usepackage{nicefrac}       
\usepackage{microtype}    \usepackage[title]{appendix}

\definecolor{bestgreen}{RGB}{0,100,0}
\definecolor{badred}{RGB}{180,0,0}

\usepackage{lipsum}
\usepackage{amsmath}
\usepackage{amsthm}
\usepackage{amssymb}
\usepackage{empheq}
\usepackage{graphicx}
\usepackage{wrapfig}
\usepackage{subcaption}
\usepackage{algorithm}
\usepackage{algpseudocode}
\usepackage{orcidlink}
\usepackage{siunitx }
\usepackage{bigints}
\usepackage{bm}
\usepackage{diagbox}
\usepackage{makecell}
\usepackage{multirow}
\usepackage{caption}
\usepackage{cleveref}
\crefname{proposition}{Proposition}{Propositions}
\crefname{lemma}{Lemma}{Lemmas}
\include{pythonlisting}
\usepackage{xcolor}
\usepackage{mathtools}
\usepackage{enumitem}
\usepackage[most]{tcolorbox}
\usepackage[numbers]{natbib}
\usepackage{hyperref}
\hypersetup{
    colorlinks=true,
    citecolor=black,
    linkcolor=black,
    filecolor=black,
    urlcolor=black,
}

\usepackage{pifont}

\definecolor{tblheaderblue}{HTML}{9CCDEA}
\definecolor{tblrowblue}{HTML}{DCEEF9}
\definecolor{tblrule}{HTML}{000000}

\newcommand{\tablestyle}{%
    \arrayrulecolor{tblrule}%
    \renewcommand{\arraystretch}{1.3}%
}

\newcommand{\RR}{\mathbb{R}}

\newtheorem{theorem}{Theorem}[section]

\newtheorem{lemma}[theorem]{Lemma}
\newtheorem{proposition}[theorem]{Proposition}

\newtheorem{remark}[theorem]{Remark}

\numberwithin{equation}{section}

\newcommand{\orcidauthor}[1]{\raisebox{0.15ex}{\scriptsize\orcidlink{#1}}}

\title{NAP-Wave: A \textbf{N}eural \textbf{A}daptive \textbf{P}lane-\textbf{Wave} Architecture for High-Frequency Helmholtz Problems}

\author{
Hamza Kamil$^{1,*}$\,\orcidauthor{0009-0009-7179-7970}
\quad
Mohammad Mahdi Abedi$^{2}$\,\orcidauthor{0000-0001-6937-0243}
\quad
David Pardo$^{2,1,3}$\,\orcidauthor{0000-0002-1101-2248}
\\
{\small
$^{1}$Basque Center for Applied Mathematics (BCAM), Bilbao, Spain
}\\
{\small
$^{2}$University of the Basque Country (UPV/EHU), Leioa, Spain
}\\
{\small
$^{3}$Ikerbasque: Basque Foundation for Science, Bilbao, Spain
}\\
{\small
\texttt{hkamil@bcamath.org}
\quad
\texttt{mohammadmahdi.abedi@ehu.eus}
\quad
\texttt{david.pardo@ehu.eus}
}
}

\begin{document}

\maketitle

\begin{abstract}
Training deep neural networks to simulate highly oscillatory wavefields presents optimization challenges, often hindered by spectral bias and poorly conditioned physics-based loss landscapes. To address these challenges and enhance convergence, we propose NAP-Wave, a joint neural architecture and training strategy that expands the target scattered wavefield in an adaptive neural plane-wave basis. The basis functions are extracted from all hidden layers and modulated by trainable plane-wave carriers, embedding the high-frequency oscillatory physics directly into the network's trial space. NAP-Wave is trained on a physics-informed loss built from the Lippmann-Schwinger integral equation, so no differentiation of the network with respect to its input is required. We develop a decoupled optimization strategy in which Adam updates the parameters of the neural plane-wave basis, while the linear combination coefficients are computed exactly by a least-squares (LS) solver. We couple this with an adaptive-depth training strategy in which the network starts shallow and grows progressively layer-by-layer. At each growth step, we avoid the loss spikes typical of dynamic network growth using the LS-computed coefficients, the warm-started parameters, and the preservation of the first- and second-order optimizer moments of the previous stage. Ablation studies confirm that the plane-wave basis, the exact coefficient solve, and the progressive schedule independently accelerate convergence and improve accuracy. Validated on exact-solution benchmarks, NAP-Wave achieves relative $L^2$ errors of order $1\%$. On highly heterogeneous geophysical benchmarks (Marmousi, Overthrust, and Otway), it achieves relative errors of $5\%$, $9\%$, and $10\%$, significantly outperforming published PINN methods in both accuracy and convergence time. To ensure reproducibility, the code is available at \url{https://github.com/Hamza-Kamil/NAP-WAVE}.
\end{abstract}

\keywords{PINNs  \and Neural basis
functions \and Plane-wave methods \and Helmholtz scattering  \and Lippmann-Schwinger equation  \and Seismic modeling}

\section{Introduction}
\label{sec:introduction}

\subsection{Background and motivation}
\label{subsec:background}

Neural PDE solvers have emerged as promising mesh-free algorithms for scientific simulations (e.g., \cite{abedi2025gabor, abedi2026least, song2021solving, nair2025multiple}). However, learning highly oscillatory, high-frequency functions remains a profound challenge for standard neural architectures. This difficulty is mainly driven by spectral bias, a phenomenon where neural networks inherently prioritize learning low-frequency components and struggle to fit high-frequency details \cite{ronen2019convergence}. Furthermore, minimizing high-frequency PDE residuals often requires computing second-order spatial derivatives, which can result in highly non-convex, poorly conditioned loss landscapes that severely impede optimizer convergence.

A particularly challenging setting for testing these neural solvers is time-harmonic wave propagation, governed by the Helmholtz equation. As the frequency and medium contrast increase, the exact solutions become highly oscillatory and multiscale. While classical discretizations require prohibitively large, indefinite linear systems to resolve these fine wavelengths, standard deep neural networks equally fail due to spectral bias and the difficulty of resolving sharp localized features in highly heterogeneous velocity models, such as those focommonly found in geophysics and medical imaging \cite{berkhout2012combining, guasch2020full}.

The Lippmann-Schwinger integral formulation eliminates second-order 
derivatives that appear in the strong form of the Helmholtz equation. As a result, the training
loss does not require differentiation of the network with respect to its input, and the associated integral
operator is better conditioned than the differential Helmholtz operator. Embedding the outgoing radiation condition into the background Green's function turns the scattering problem into a Fredholm integral equation of the second kind. The differential Helmholtz operator is unbounded and ill-conditioned, whereas the resulting integral operator is a bounded identity-plus-compact operator. This structural preconditioning gives a smoother loss landscape and helps stabilize training. The action of the Green operator on the trial space is evaluated on a uniform grid via FFT, exploiting the Toeplitz structure of the kernel. This evaluation is far faster than direct evaluation of the Helmholtz operator, which requires differentiating the network at every collocation point.

Abedi et al. \cite{abedi2026green} proposed the Lippmann-Schwinger integral equation  as a training loss for neural solvers of scattering problems, combining it with an auxiliary strong-form residual as a regularizer. We adopt the same Green-integral loss and instead improve the architecture and optimization strategy: for the same loss, our design converges faster and reaches lower error in strongly heterogeneous media. The NN architecture of the trial space is designed so that the wave structure is represented directly by the network, and it is paired with an optimization strategy built specifically to make training fast and stable.

In this study, we introduce NAP-Wave, a data-free neural solver built on three design principles.

\begin{enumerate}
\item \textbf{The wave structure belongs in an adaptive plane-wave trial space.} A generic network must discover, from a smooth coordinate input, a field oscillating on the scale of a wavelength. Trefftz and plane-wave methods avoid this by building basis functions that satisfy the governing equation exactly \cite{imbert2014generalized, colton2001denseness}. We adopt the same principle in a neural setting. Each basis function carries a plane-wave that solves the homogeneous Helmholtz equation exactly, and the network represents the varying amplitude that modulates it.

\item \textbf{The linear coefficients are solved exactly, while the basis adapts.} The Lippmann-Schwinger operator is linear, so the coefficients combining the basis functions admit a closed-form solution for any fixed basis. We solve for them by least squares at every step. Adam updates only the nonlinear parameters that shape the basis functions themselves. This is the classical variable projection principle \cite{golub1973differentiation}, applied here with every neuron of every hidden layer treated as an individual basis function.

\item \textbf{The trial space grows with training, rather than staying fixed.} A network initialized at full depth must resolve a high-wavenumber field from a random basis. We instead train a sequence of progressively deeper sub-networks. Each new layer adds a block of basis functions and preserves the first- and second-order moments of the optimizer from the converged parameters of the previous stage, so the trial space is enriched only once the coarser one has been exploited. Networks are known to learn low-frequency components first \cite{ronen2019convergence}; growing depth progressively lets each new layer target the higher-frequency content the shallower network could not yet represent. In practice, we find that this progressive schedule acts as a warm start: once a shallow sub-network has converged, its parameters already place the next, deeper sub-network in a good basis, and adding further training iterations to the shallow network alone yields no additional gain.
\end{enumerate}

\subsection{Related work}
\label{subsec:related_work}

\paragraph{Integral-equation solvers for scattering.}
The Lippmann-Schwinger equation is a classical tool in scattering theory. It appears across many fields, including quantum mechanics \cite{ryoo2025lippmann}, electromagnetics \cite{costabel2012essential}, acoustics \cite{lechleiter2015time}, and seismic wave propagation \cite{jakobsen2020homotopy}, and we adopt it here for frequency-domain seismic scattering. Its mathematical foundations
are well established, including compactness of the volume potential,
Fredholm invertibility, and radiation conditions \cite{kirsch2011introduction}.
Traditional numerical solvers discretize the integral operator and assemble
the resulting dense system. This system is prohibitively expensive to
invert directly. Practitioners therefore use iterative solvers instead, in
which each matrix-vector product is evaluated efficiently via the FFT. This
approach exploits the Toeplitz structure of the Green's kernel on uniform
grids. The Born series \cite{morse1946methods} is a classical instance of
this iterative approach. It converges for weak scattering contrasts, but
strong contrasts require effective preconditioners to obtain a convergent
iteration \cite{10.1093/gji/ggaa503,osnabrugge2016convergent}. The FFT-accelerated evaluation of the Green operator that we adopt in this
study relies on the same classical construction where we use it to
assemble the action of the operator on the trial space spanned by the
plane-wave neural basis functions. This yields a differentiable solver for the scattered field that can be evaluated at
arbitrary spatial locations.

\paragraph{Trefftz methods, plane-wave basis, and Herglotz wave functions.}
Trefftz and quasi-Trefftz methods use basis functions that satisfy the
governing equation exactly, or up to a high-order local error, rather than
generic polynomials. Generalized plane-wave methods extend this idea to
smoothly varying media \cite{imbert2014generalized}. Our plane-wave neural
basis is related in spirit: the plane-wave carriers solve the homogeneous
Helmholtz equation exactly, while the neural amplitude captures the spatially
varying correction. Our approach differs from plane-wave activation networks
\cite{cui2022efficient}, in which plane-waves are used as the activation
functions of a single-hidden-layer network, whereas we combine plane-wave
carriers with a deep neural amplitude and target general heterogeneous,
spatially varying wavenumber media rather than a homogeneous background. It
also differs from the plane-wave discontinuous Galerkin domain-decomposition
method of \cite{yuan2025discontinuous}, which is formulated for
piecewise-constant wavenumbers, whereas our method addresses smoothly and
strongly varying velocity models within a single global trial space. 

Our approach is based on Herglotz wave functions, which superpose plane-waves at a fixed
wavenumber and integrated over all directions on the unit sphere. Classical results show such
superpositions are dense in the solution space of the homogeneous Helmholtz equation, with
approximation properties in Sobolev norms studied rigorously \cite{colton2001denseness}. Each basis function carries a plane wave at a fixed direction, sampled from a finite set of directions on the unit sphere, together with a trainable wavenumber and a trained amplitude that modulates it. The homogeneous solution space is thus approximated over a finite, discrete set of directions by varying wavenumber and amplitude, rather than over a continuum of directions at a single fixed wavenumber. Classical density results for plane-wave bases, such as \cite{colton2001denseness}, are stated for a continuum of directions at fixed wavenumber, so they do not apply to our construction. We prove a separate density result for our finite-direction, trainable-wavenumber basis in Theorem~\ref{th:universal_approximation}.

\paragraph{Neural solvers for Helmholtz and seismic modeling.}
PINNs and related neural approaches have been
applied to frequency-domain wave modeling in heterogeneous media, including
Gabor, Fourier-feature, and Green-integral-constrained formulations
\cite{abedi2025gabor, abedi2026green, abedi2026least, ding2025physics,
song2021solving}. These methods train the network end-to-end on the Helmholtz equation or its integral
formulation, relying solely on the network's own representational capacity, without introducing an
inductive bias such as the plane-wave basis proposed here. Oscillatory structures are generally incorporated into the network architecture through input encoding techniques, such as positional encoding \cite{abedi2025gabor}, or by warm-starting the training from models previously trained at lower frequencies \cite{huang2022pinnup}.  In contrast, our approach embeds this structure directly in the trial space through the plane-wave carriers, which operate effectively on the physical coordinates without any feature embedding.

\paragraph{Least-squares for the linear coefficients.}
The idea of adaptive basis functions with a linear output layer solved by least squares appears independently in several lines of work. Extreme learning machines fix a single hidden layer at random and solve only for the output weights by least squares \cite{huang2006extreme}. The variable projection (VarPro) principle for nonlinear least-squares problems formalizes this split more generally: the linear parameters are solved using least squares, and the first-order optimization is performed over the nonlinear parameters alone \cite{golub1973differentiation}. This principle has been adopted in PINNs (see e.g., \cite{abedi2026least, baharlouei2026least}), where the network parameters are optimized by Adam while the weights of the last layer are solved exactly by least squares. Our approach differs in that all neural basis functions, taken from every hidden layer rather than only the last, are connected to the output, treating each neuron as an individual basis function. We then use the exact least-squares solve for the coefficients coupled with gradient-based training of the nonlinear network parameters. Unlike a single hidden layer, a deep network builds hierarchical features across layers, so treating every hidden neuron as a basis function lets us capture local, fine-scale structure in the field that a shallow basis cannot represent.

\subsection{Contributions and paper organization}
\label{subsec:contributions}
We summarize our contributions as follows.
\begin{itemize}
\item A plane-wave neural basis for time-harmonic wave propagation, in which every neuron of every hidden layer contributes a basis function modulated by an oscillatory plane-wave carrier with trainable wavenumber and amplitude.
\item An extension of the variable projection scheme to all layers, in which the linear coefficients are obtained at each step from a regularized least-squares solve, while gradient-based training updates only the nonlinear carrier and amplitude parameters.
\item An adaptive-depth training strategy that grows the architecture layer-by-layer instead of fixing it at full depth from the start. Combined with the exact coefficient solve, it removes the loss spikes that occur when a new layer is added and keeps the error decreasing smoothly, an effect that neither ingredient produces alone.
\item A GPU implementation in JAX combining an FFT evaluation of the Green operator with a complex-valued direct solve in double precision. It reaches relative $L^2$ errors of order $1\%$ against exact 2D and 3D solutions, and $5\%$, $9\%$, and $10\%$ against the reference solution on Marmousi (10~Hz), Overthrust (10~Hz), and Otway (20~Hz), with the 10-Hz Marmousi case solved in roughly three minutes on a single GPU.
\end{itemize}

The remainder of the paper is organized as follows. In Section~\ref{sec:mathematical_models}, we recall the Lippmann-Schwinger equation and its abstract properties. In Section~\ref{sec:methods}, we present the proposed methodology. In Section~\ref{sec:implementation}, we describe the algorithmic and computational aspects of the proposed method. In Section~\ref{sec:case_studies}, we describe the benchmark problems and the reference solutions. In Section~\ref{sec:numerical_results}, we report the numerical experiments. In Section~\ref{sec:conclusion}, we summarize the main conclusions of this study. Further details are given in Appendices~\ref{appendix:arch}-\ref{app:fft_implementation}.

\section{Mathematical model: The Lippmann-Schwinger equation}
\label{sec:mathematical_models}

In this section, we recall the Lippmann-Schwinger equation, the mathematical model adopted in
this study to describe time-harmonic acoustic wave propagation in heterogeneous media.

\subsection{Setting and medium contrast}
Let $\Omega \subset \mathbb{R}^d$ be a bounded connected open set with $d = 2, 3$. We consider
time-harmonic wave propagation at angular frequency $\omega = 2\pi f_0 > 0$, where $f_0 > 0$
denotes the temporal frequency. The wave speed is denoted by $v : \Omega \to \mathbb{R}$ and we
assume that there exist constants
$0 < v_{\min} \le v_{\max} < \infty$ such that $v_{\min} \le v(x) \le v_{\max}$ for all
$x \in \Omega$. The local wavenumber is given by $k(x) = \omega / v(x)$, and we denote the squared
slowness by $m(x) = v(x)^{-2}$. The corresponding background squared slowness and wavenumber are
$m_0 = v_0^{-2}$ and $k_0 = \omega / v_0$, respectively, where $v_0 > 0$ is the constant background
wave speed. The medium contrast, which quantifies the deviation of the true medium from the background, is
defined by
\begin{equation}
    \delta m(x) = m(x) - m_0,
\end{equation}
and satisfies $\delta m \in L^\infty(\Omega)$ with $\operatorname{supp}(\delta m) \subset \Omega$,
which ensures that the scatterer is compactly supported within $\Omega$.

\subsection{Incident and scattered fields}
The total acoustic field $u$ is decomposed as
\begin{equation}
    u = u_0 + u_s,
\end{equation}
where $u_0$ denotes the incident field, propagating in a prescribed homogeneous background medium
with constant wave speed $v_0 > 0$, and $u_s$ is the scattered field arising from the medium
heterogeneities.

Let $G_0$ denote the outgoing free-space Green's function of the background Helmholtz operator
$\mathcal{L}_0 := -\Delta - \omega^2 m_0$, that is, the unique solution of
$\mathcal{L}_0 G_0(\cdot, y) = \delta_y$ satisfying the Sommerfeld radiation condition at infinity.
Since $v_0$ is constant, $G_0$ is translation invariant and admits the explicit kernel
representation
\begin{equation}
\label{eq:green_kernel}
    G_0(x, y) = g_0(x - y), \qquad x \neq y.
\end{equation}
In this study, we consider the free-space Green's function
\begin{equation}
\label{eq:nD_free_space_kernel}
  g_0(x)= \begin{cases}  
  \dfrac{i}{4} H_0^{(2)}\!\bigl(k_0 \|x\|\bigr), \qquad d = 2,
\\[8pt]
    \dfrac{\exp\bigl(-i k_0 \|x\|\bigr)}{4\pi \|x\|}, \qquad d = 3,
    \end{cases}   
\end{equation}
where $H_0^{(2)}$ is the Hankel function of the second kind and order zero. 
For a point source
located at $x_{\mathrm{s}} \in  \Omega$, the incident field is accordingly the
outgoing solution of the homogeneous background problem,
\begin{equation}
\label{eq:incident_field}
    u_0(x) = g_0(x - x_{\mathrm{s}}).
\end{equation}

\subsection{The Green's integral operator and the governing equation}
The Green's integral operator $\mathcal{G} : L^2(\Omega) \to L^2(\Omega)$ is defined, for
$\varphi \in L^2(\Omega)$ and $x \in \Omega$, by
\begin{equation}
\label{eq:green_operator}
    (\mathcal{G}\varphi)(x) := \omega^2 \int_\Omega g_0(x - y)\, \delta m(y)\, \varphi(y) \, \mathrm{d}y.
\end{equation}
The scattered field $u_s$ we seek must satisfy the Lippmann-Schwinger equation: given an incident
field $u_0 \in L^2(\Omega)$, find $u_s \in L^2(\Omega)$ such that
\begin{equation}
\label{eq:lippmann_schwinger}
    (I - \mathcal{G})\, u_s = \mathcal{G} u_0,
\end{equation}
where $I$ denotes the identity operator on $L^2(\Omega)$.

\subsection{Abstract formulation of the Lippmann-Schwinger equation}

This section outlines the abstract formulation of the Lippmann-Schwinger
equation. These theoretical foundations are well-established within scattering
theory, and a comprehensive mathematical treatment is available
in~\cite{kirsch2011introduction}.

The Green's operator defined in~\eqref{eq:green_operator} is a linear map
$\mathcal{G} : L^2(\Omega) \to L^2(\Omega)$ that possesses several mathematical
properties under suitable, physically realistic assumptions. First,
Lemma~\ref{lemma:K_is_Hilbert_Schmidt} (Appendix~\ref{sec:theoretical_results})
establishes that $\mathcal{G}$ is a Hilbert-Schmidt operator, which in particular
implies that it is compact and bounded. Moreover, $\mathcal{G}$ maps $L^2(\Omega)$
into $H^2_{\mathrm{loc}}(\RR^d)$, where the local Sobolev space is defined by
\begin{equation}
\label{eq:Hloc}
    H^2_{\mathrm{loc}}(\RR^d)
    =
    \bigl\{
        u : \RR^d \to \mathbb{C}
        \;\big|\;
        u|_{K} \in H^2(K)
        \text{ for every bounded open set } K \subset \RR^d
    \bigr\}.
\end{equation}
Consequently, $\mathcal{G}$ acts as a smoothing operator that elevates the spatial
regularity by two orders of differentiation.

The Lippmann-Schwinger equation~\eqref{eq:lippmann_schwinger} is formulated as
follows. Given a source term $f=\mathcal Gu_0 \in L^2(\Omega)$, we seek the unique
scattered field $u_s \in L^2(\Omega)$ satisfying
\begin{equation}
\label{eq:abstract_LS}
    \mathcal{B} u_s = f,
\end{equation}
where $\mathcal{B} = \mathcal{I} - \mathcal{G}$. The following proposition shows that  $\mathcal{B}$ is boundedly invertible on
$L^2(\Omega)$,  its proof, which
follows the classical Fredholm-alternative and unique-continuation argument of
\cite{kirsch2011introduction}, is given in Appendix~\ref{proof:invertibility_B}.

\begin{proposition}[Well-posedness of the scattering operator]
\label{prop:invertibility_B}
Let $\Omega \subset \RR^d$ be a bounded connected open set with $d=2, 3$, and let
$\delta m \in L^\infty(\Omega)$. Let $G_0(x,y)$ be the outgoing free-space
Green's function of the background operator
$L_0 = -\Delta - \omega^2 m_0$, where $k_0 = \omega/v_0$. Then the operator
$\mathcal{B} = \mathcal{I} - \mathcal{G} : L^2(\Omega) \to L^2(\Omega)$
is boundedly invertible.
\end{proposition}

\section{Approximation space and training methodology}
\label{sec:methods}

In this study, our main objective is the accurate approximation of the scattered wavefield $u_s$ in
heterogeneous media for realistic benchmark scenarios. We develop a data-free numerical approach
that employs a NN ansatz to approximate the unknown field $u_s$, trained in an
unsupervised manner directly by enforcing the Lippmann-Schwinger equation
\eqref{eq:lippmann_schwinger}.

\subsection{Residual minimization}

Since closed-form exact solutions of heterogeneous media scattering problems are
generally unavailable, we seek an approximate solution via a residual minimization
method. Recall that the scattered field $u_s \in L^2(\Omega)$ satisfies the
Lippmann-Schwinger equation $\mathcal{B}u_s = f$, where $\mathcal{B} = \mathcal{I} -
\mathcal{G}$ and $\mathcal{G}$ is the integral operator defined in
\eqref{eq:green_operator}. Given an incident source $f \in L^2(\Omega)$, we introduce
the least-squares functional $J : L^2(\Omega) \to \RR_{\geq 0}$ defined by
\begin{equation}
\label{eq:cost_function_general}
    J(v) = \frac{1}{2}\|\mathcal{B}v - f\|_{L^2(\Omega)}^2.
\end{equation}
As shown in Theorem~\ref{thm:existence_uniqueness_minimizer}, $J$ is a strictly
convex quadratic functional admitting a unique global minimizer, which coincides
with the unique solution of the governing integral equation; the proof is given in
Appendix~\ref{proof:existence_uniqueness_minimizer}.

\begin{theorem}[Existence and uniqueness of the global minimizer]
\label{thm:existence_uniqueness_minimizer}
The least-squares functional $J : L^2(\Omega) \to \RR_{\geq 0}$ given by \eqref{eq:cost_function_general} 
is strictly convex and quadratic, and admits a unique global minimizer
$u_s \in L^2(\Omega)$, which is the unique solution of $\mathcal{B}u_s = f$.
\end{theorem}

A central question is whether minimizing this residual-based loss also controls
the true approximation error, e.g., in the $L^2$ sense. We show that this is
indeed the case and therefore,
minimizing $J$ is equivalent to minimizing the true error, as stated precisely in
the following theorem; see Appendix~\ref{proof:error_estimate} for the proof.

\begin{theorem}[Error estimate via residual control]
\label{thm:error_estimate}
Let $f \in L^2(\Omega)$, let $u \in L^2(\Omega)$ be the exact solution of
$\mathcal{B}u = f$, and let $v \in L^2(\Omega)$ be an arbitrary approximation.
Define the approximation error $e(v) = u - v$. Then, there exist constants $c_1, c_2 > 0$ such that
\begin{equation}
\label{eq:loss_error_bound}
    c_1 \sqrt{J(v)}
    \;\le\;
    \frac{\|e(v)\|_{L^2(\Omega)}}{\|u\|_{L^2(\Omega)}}
    \;\le\;
    c_2 \sqrt{J(v)}.
\end{equation}
\end{theorem}

The ratio of the equivalence constants in~\eqref{eq:loss_error_bound} is given by the
condition number $\kappa(\mathcal{B}) = c_2/c_1= \|\mathcal{B}\|\,\|\mathcal{B}^{-1}\|$.
In the high-frequency or strong-contrast regime, $\|\mathcal{G}\|$ grows and
$\mathcal{B} = \mathcal{I} - \mathcal{G}$ becomes ill-conditioned, so a small
residual no longer guarantees a small error.

We now restrict the minimization of $J$ to a
finite-dimensional trial space constructed from NNs and seek an
approximation of the scattered field $u_s$ within it.

\subsection{Approximation space}
\label{sec:approximation_space}

We seek to approximate the unknown scattered field $u_s \in L^2(\Omega)$ in a
space spanned by neural basis functions. For a given feedforward NN,
each neuron in each hidden layer defines a real-valued function of $x \in
\Omega$, and consecutive neuron pairs yield complex-valued basis functions. A schematic
overview is given in Figure~\ref{fig:basis_construction} and the
full construction is given in Appendix~\ref{appendix:arch}.

Let $\theta \in \mathbb{R}^p$ denote the vector of network parameters, with $p$ the total number of parameters. For a fixed $\theta$, let $\varphi_j(\cdot\,;\theta) : \Omega \to \mathbb{C}$ for $j = 1, \ldots, N$ be complex-valued functions generated by a NN. We consider the approximation space
\begin{equation}
\label{eq:VN_space}
    V_N(\theta)
    =
    \operatorname{span}
    \left\{
        \varphi_1(\cdot\,;\theta), \ldots, \varphi_N(\cdot\,;\theta)
    \right\},
\end{equation}
and we seek $u_N \in V_N(\theta)$ of the form
\begin{equation}
\label{eq:approximation}
    u_N(x;\theta,\bm{c})
    =
    \sum_{j=1}^{N} c_j\,\varphi_j(x;\theta),
    \qquad x \in \Omega,
\end{equation}
where $\bm{c} = (c_j)_{j=1}^N \in \mathbb{C}^N$.

The nonlinear network parameters $\theta$ and the linear coefficients $\bm{c}$
are determined by minimizing $J$ defined in~\eqref{eq:cost_function_general}, as
we detail in Section~\ref{sec:Optimisation}. Note that only $\theta$ requires
initialization, and the coefficients $\bm{c}$ need not be initialized, since for any
given $\theta$ they are determined in closed form. Moreover, since $\theta$ is updated during
training, the space $V_N(\theta)$ itself adapts, and the method constitutes an
adaptive basis strategy. For a fixed number of basis functions $N$, the search
space over all admissible parameters is denoted as
\begin{equation}
    \mathcal{V}_{N}
    =
    \bigcup_{\theta \in \RR^p} V_N(\theta).
\end{equation}

We note that the functions $\varphi_j$ are not required to be linearly independent
in general. In this study, we consider two types of neural basis functions.

\textbf{Standard neural basis functions.} The basis functions are formed
directly from hidden neuron pairs, without any additional structure imposed
on the network; see~\eqref{eq:complex_basis_all_layers}. This construction
relies entirely on training to discover any oscillatory 
behavior present in the solution.

\textbf{Plane-wave neural basis functions.} Each basis function is formed as the product of a hidden neuron pair and an
oscillatory carrier $\exp(\mathrm{i} k_j \bm{d}_j \cdot x)$, where the
propagation direction $\bm{d}_j$ is sampled once at initialization and held
fixed, while the wavenumber $k_j$ is a trainable network parameter determined
during training, as described in \eqref{eq:plane_wave_basis}. By
construction, each carrier is an exact solution of the homogeneous Helmholtz
equation, so that the dominant oscillatory structure of $u_s$ is built
directly into the basis rather than left for the network to learn from
scratch. The remaining task for the network is then to represent the
spatially varying complex amplitude that modulates each carrier, which is
typically a much smoother function than $u_s$ itself and could therefore be
easier to approximate.

Note that plane-wave basis functions form a generalization of the standard basis; 
setting the wave-number samples $k_j = 0$ for all $j$, and the standard basis is recovered
as a special case.

\begin{figure}[!h]
    \centering
    \begin{subfigure}{\linewidth}
        \centering
               \caption{Neural basis function construction}
    \includegraphics[width=\linewidth]{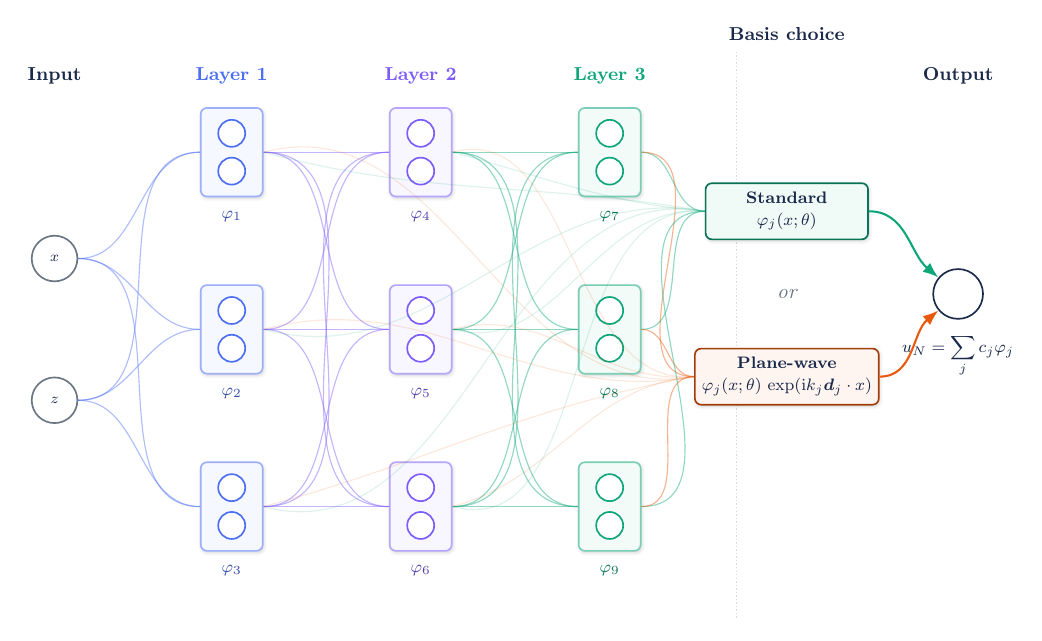}
        \label{fig:basis_construction}
    \end{subfigure}
    \begin{subfigure}{\linewidth}
        \centering
           \caption{Plane-wave neuron construction}
    \includegraphics[width=\linewidth]{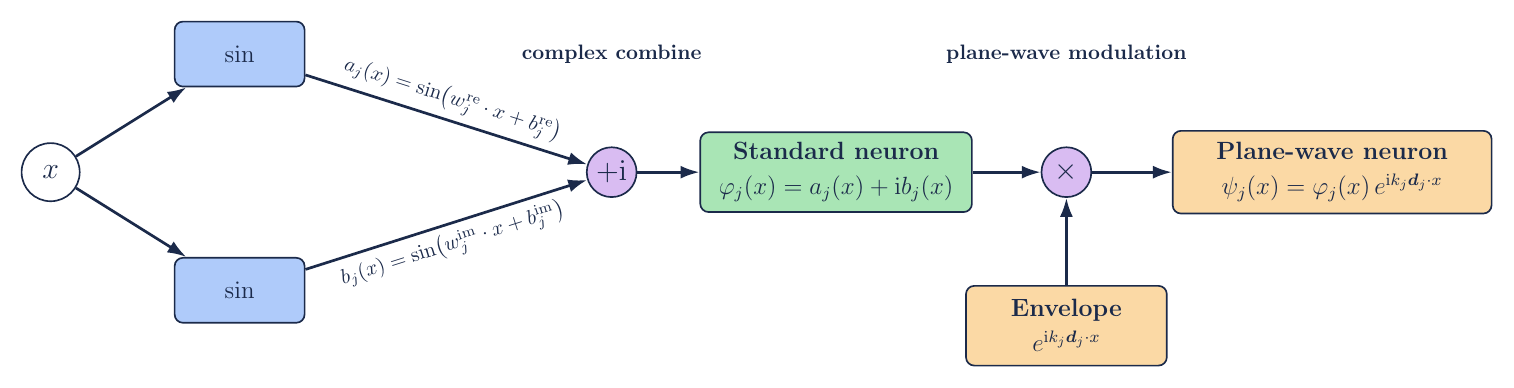}
        \label{fig:plane-wave-neuron}
    \end{subfigure}
    \caption{Construction of the neural basis functions from a feedforward
    network architecture, illustrated for a network with three hidden
    layers. (a)~Consecutive neurons within each layer are paired to form
    complex-valued basis functions; 
    each hidden layer therefore has an even number of neurons. Basis
    functions from every layer are connected directly to the output, so the two constructions shown, \emph{standard}
    (using the complex pairs directly) and \emph{plane-wave} (multiplying each
    pair by an oscillatory carrier $\exp(\mathrm{i}k_j\bm{d}_j\cdot x)$). (b)~Detailed construction of a
    single plane-wave neuron: two real sub-neurons produce the real and
    imaginary parts $a_j,b_j$, which combine into the complex feature
    $\varphi_j = a_j + \mathrm{i}b_j$; this is then modulated by the carrier
    $\exp(\mathrm{i}k_j\bm{d}_j\cdot x)$ to yield the plane-wave neuron $\psi_j$.}
    \label{fig:basis_construction}
\end{figure}

In the following, we show that the considered approximation space is dense in $L^2(\Omega)$, a natural result stemming from the density of NNs reformulated as an adaptive basis strategy. We denote by $\mathcal{V}$ the class generated by the standard basis and by
$\mathcal{V}_w$ the class generated by the plane-wave basis, each taken over all
possible numbers of basis functions $N$, that is
$\mathcal{V} = \bigcup_{N} \mathcal{V}_N$. Since the solution of the
Lippmann-Schwinger equation belongs to $L^2(\Omega)$, we aim to show that both
$\mathcal{V}$ and $\mathcal{V}_w$ are dense in $L^2(\Omega)$.
\begin{theorem}[Universal approximation]
\label{th:universal_approximation}
Let $\Omega \subset \mathbb{R}^d$ be a bounded Lipschitz domain, and let $\sigma$ be a continuous, non-polynomial activation function. Then the neural approximation classes $\mathcal{V}$ and $\mathcal{V}_w$ are dense in $L^p(\Omega)$ for all $p \in [1, \infty)$.
\end{theorem}
\begin{proof}
See Appendix~\ref{proof:universal_approximation}.
\end{proof}

\subsection{Optimization}
\label{sec:Optimisation}

We now present the optimization procedure adopted to determine $\theta$ and
$\bm{c}$. For a fixed number of basis functions $N$, the network parameters
$\theta \in \mathbb{R}^p$ define the approximation space $V_N(\theta)$.
Within this space, the optimal coefficients $\bm{c}$ are recovered by
minimizing $J$ over $\bm{c}$ alone; since $J$ is quadratic in $\bm{c}$ for
fixed $\theta$, this subproblem reduces to a linear least-squares problem
with a closed-form solution. The parameters $\theta$ are then updated to
refine the space $V_N(\theta)$, and the two steps alternate until a
convergence criterion is met. This variable projection strategy decouples
the linear and nonlinear parts of the optimization, and we give more details
in the following sections.

\subsubsection{Construction of the loss function on $\mathcal{V}_N$}

We seek the best approximation of $u_s$ within $\mathcal{V}_N$, which amounts
to finding optimal parameters $(\theta, \bm{c}) \in \RR^p \times
\mathbb{C}^N$ minimizing the least-squares functional $J$ defined
in~\eqref{eq:cost_function_general}. For fixed $\theta \in \RR^p$, any
$v \in V_N(\theta)$ takes the form
\begin{equation}
\label{eq:v_expansion}
    v(x;\theta,\bm{c})
    =
    \sum_{j=1}^{N} c_j\,\varphi_j(x;\theta),
    \qquad x \in \Omega,\quad c_j \in \mathbb{C}.
\end{equation}
By linearity of $\mathcal{B}$, the functional $J$ evaluated at $v$ reads
\begin{equation}
\label{eq:J_discrete}
    J(v)
    =
    \frac{1}{2}
    \left\|
        \sum_{j=1}^{N} c_j\,\mathcal{B}\varphi_j(\cdot\,;\theta) - f
    \right\|_{L^2(\Omega)}^2.
\end{equation}
To isolate the linear structure in $\bm{c}$, we define the operator
$\Phi_\theta : \mathbb{C}^N \to V_N(\theta)$ by
\begin{equation}
\label{eq:Phi_theta}
    \Phi_\theta\,\bm{c}
    :=
    \sum_{j=1}^{N} c_j\,\varphi_j(\cdot\,;\theta),
\end{equation}
and $J$ takes the compact form
\begin{equation}
\label{eq:J_compact}
    J(\theta, \bm{c})
    =
    \frac{1}{2}\big\|\mathcal{B}\Phi_\theta\,\bm{c} - f\big\|_{L^2(\Omega)}^2.
\end{equation}
We aim to solve the optimization problem
\begin{equation}
\label{eq:full_opt}
    (\theta^*, \bm{c}^*)
    \in
    \operatorname*{arg\,min}_{\theta \in \RR^p,\; \bm{c} \in \mathbb{C}^N}
    J(\theta, \bm{c}),
\end{equation}
which is equivalently reformulated as the nested problem
\begin{equation}
\label{eq:reduced_opt}
    \min_{\theta \in \RR^p}
    \left(
    \min_{\bm{c} \in \mathbb{C}^N}
    J(\theta, \bm{c})
    \right).
\end{equation}
This formulation separates the joint problem into two subproblems:
for fixed $\theta$, the inner minimization over $\bm{c}$ is a linear
least-squares problem, and the outer minimization over $\theta$ is a nonlinear
problem addressed using gradient-based optimization.

\subsubsection{Determination of the linear coefficients $\mathbf{c}$}

For fixed $\theta$, the map $\bm{c} \mapsto J(\theta, \bm{c})$ is quadratic,
and the optimal coefficients are defined by
\begin{equation}
\label{eq:c_opt}
    \bm{c}(\theta)
    \in
    \operatorname*{arg\,min}_{\bm{c} \in \mathbb{C}^N}
    J(\theta, \bm c).
\end{equation}
In general, problem~\eqref{eq:c_opt} may admit multiple solutions, and we
therefore replace it by the well-posed Tikhonov-regularized problem
\begin{equation}
\label{eq:tikhonov}
    \bm{c}_\mu(\theta)
    =
    \operatorname*{arg\,min}_{\bm{c} \in \mathbb{C}^N}
    \left\{
        J(\theta, \bm c)
        +
        \frac{\mu}{2}\|\bm{c}\|_{\mathbb{C}^N}^2
    \right\},
\end{equation}
where $\mu > 0$ is the regularization parameter. Since $\mathcal{B}\Phi_\theta :
\mathbb{C}^N \to L^2(\Omega)$ is bounded and its adjoint
$(\mathcal{B}\Phi_\theta)^* : L^2(\Omega) \to \mathbb{C}^N$ is well defined,
problem~\eqref{eq:tikhonov} is equivalent to solving the normal equations
\begin{equation}
\label{eq:normal_equations}
    \bigg((\mathcal{B}\Phi_\theta)^* \mathcal{B}\Phi_\theta + \mu I\bigg)\bm{c}_\mu(\theta) = (\mathcal{B}\Phi_\theta)^* f.
\end{equation}
The normal equations~\eqref{eq:normal_equations} have a unique solution for each
$\theta \in \RR^p$ and $\mu > 0$. The regularity of $\bm{c}_\mu$ as a function of
$\theta$ is inherited from the regularity of the basis functions, and the
following proposition establishes that $\theta \mapsto \bm{c}_\mu(\theta)$ is
continuously differentiable.

\begin{proposition}[Regularity of $\bm{c}_\mu$]
\label{prop:c_theta_continuous}
Assume that for each $j = 1, \ldots, N$, the basis function
$\varphi_j(\cdot\,;\theta)$ is continuously differentiable with respect to
$\theta \in \RR^p$. Then,  for each $\mu > 0$, the map
\begin{equation}
    \bm{c}_\mu : \RR^p \to \mathbb{C}^N,
    \qquad
    \theta \mapsto \bm{c}_\mu(\theta)
    =
    \left((\mathcal{B}\Phi_\theta)^* \mathcal{B}\Phi_\theta + \mu I\right)^{-1} (\mathcal{B}\Phi_\theta)^* f,
\end{equation}
is well defined and continuously differentiable.
\end{proposition}
\begin{proof}
See Appendix~\ref{proof:c_theta_continuous}.
\end{proof}

\subsubsection{Determination of the nonlinear parameters $\theta$}

Substitution of $\bm{c}_\mu(\theta)$ into~\eqref{eq:J_compact} yields the
reduced loss
\begin{equation}
\label{eq:reduced_loss}
    \mathcal{J}_\mu(\theta)
    :=
    J(\theta, \bm{c}_\mu(\theta))
    =
    \frac{1}{2}\left\|\mathcal{B}\Phi_\theta\bm{c}_\mu(\theta) - f\right\|_{L^2(\Omega)}^2,
\end{equation}
which depends on $\theta$ alone. The outer minimization
in~\eqref{eq:reduced_opt} thus reduces to
\begin{equation}
\label{eq:outer_opt}
    \theta^*
    \in
    \operatorname*{arg\,min}_{\theta \in \RR^p}
    \mathcal{J}_\mu(\theta).
\end{equation}
A key practical advantage of this formulation is that differentiation through
the map $\theta \mapsto \bm{c}_\mu(\theta)$ is not required. Indeed, the following
proposition shows that the gradient of the reduced loss equals the partial
gradient of the joint loss with respect to $\theta$, evaluated at the optimal
coefficients.

\begin{proposition}[Reduced gradient identity]
\label{prop:gradient_reduced_loss}
Let $\mu > 0$ and assume that each $\varphi_j(\cdot\,;\theta)$ is continuously
differentiable in $\theta$. Define the regularized loss
\begin{equation}
\label{eq:joint_loss_methods}
    J_\mu(\theta, \bm{c})
    :=
    J(\theta, \bm{c})
    +
    \frac{\mu}{2}\|\bm{c}\|_{\mathbb{C}^N}^2.
\end{equation}
Then the reduced loss $\mathcal{J}_\mu(\theta) := J_\mu(\theta,
\bm{c}_\mu(\theta))$ is differentiable and
\begin{equation}
\label{eq:reduced_gradient}
    \nabla \mathcal{J}_\mu(\theta)
    =
    \nabla_\theta J(\theta, \bm{c}_\mu(\theta)).
\end{equation}
\end{proposition}
\begin{proof}
See Appendix~\ref{proof:gradient_reduced_loss}.
\end{proof}

To update the nonlinear parameters of the basis functions, we use
Adam~\cite{kingma2015adam} to solve the outer nonlinear optimization
problem~\eqref{eq:outer_opt}. Adam is a stochastic first-order optimizer
that preconditions the gradient descent update with a diagonal approximation
of the curvature, assigning each parameter an adaptive learning rate from
exponential moving averages of the first and second moments of the gradient.
It is the default choice in the machine learning community for its
scalability and efficiency in high dimensions. Given
$g_n = \nabla_\theta \mathcal{J}_\mu(\theta_n)$, the update reads
\begin{equation}
     \theta_{n+1} = \theta_n - \eta_n\frac{\hat{m}_n}{\sqrt{\hat{v}_n}+\varepsilon},
\end{equation}
where
\begin{equation}
    m_n = \beta_1 m_{n-1} + (1-\beta_1)g_n,
    \qquad
    v_n = \beta_2 v_{n-1} + (1-\beta_2)g_n \odot g_n,
\end{equation}
\begin{equation}
    \hat{m}_n = \frac{m_n}{1-\beta_1^n},
    \qquad
    \hat{v}_n = \frac{v_n}{1-\beta_2^n},
\end{equation}
with default hyperparameters $(\beta_1, \beta_2, \varepsilon) = (0.9, 0.999, 10^{-8})$ and the
learning rate is decayed by a factor of $0.9$ every $2000$ steps.

\subsection{Discretization of the normal equations and the empirical loss}
The normal equations~\eqref{eq:normal_equations} and the reduced loss~\eqref{eq:reduced_loss}
involve $L^2(\Omega)$ inner products that must be approximated numerically. Let
$\{x_k\}_{k=1}^{N_q} \subset \Omega$ denote quadrature points forming a uniform grid with
equal weights $|\Omega|/N_q$. Approximating the $L^2(\Omega)$ inner products by this quadrature
rule turns the continuous normal equations~\eqref{eq:normal_equations} into the discrete normal
equations
\begin{equation}
\label{eq:discrete_normal_equations}
    \big(\mathsf{B}_\theta^* \mathsf{B}_\theta + \mu I\big)\,
    \bm{c}^{(N_q)}_\mu(\theta)
    =
    \mathsf{B}_\theta^* \bm{f},
\end{equation}
where the matrix and vector entries are given by
\begin{equation}
\label{eq:discrete_entries}
    \big(\mathsf{B}_\theta^* \mathsf{B}_\theta\big)_{ij}
    :=
    \sum_{k=1}^{N_q}
    \mathcal{B}\varphi_i(x_k;\theta)\,
    \overline{\mathcal{B}\varphi_j(x_k;\theta)},
    \qquad
    \big(\mathsf{B}_\theta^* \bm{f}\big)_i
    :=
    \sum_{k=1}^{N_q}
    f(x_k)\,
    \overline{\mathcal{B}\varphi_i(x_k;\theta)},
\end{equation}
with $\mathsf{B}_\theta \in \mathbb{C}^{N_q \times N}$ defined entrywise by
$(\mathsf{B}_\theta)_{kj} = \mathcal{B}\varphi_j(x_k;\theta)$, adjoint
$\mathsf{B}_\theta^* = \overline{\mathsf{B}_\theta}^\top$, and
$\bm{f} = (f(x_1),\dots,f(x_{N_q}))^\top$. 

Correspondingly, the empirical green-integral (GI) loss takes the vector form
\begin{equation}
\label{eq:empirical_loss_vector}
    \mathcal{J}^{(N_q)}_\mu(\theta)
    =
    \frac{|\Omega|}{2N_q}
    \big\|
        \mathsf{B}_\theta\,\bm{c}^{(N_q)}_\mu(\theta) - \bm{f}
    \big\|_2^2
    \propto
    \frac{1}{N_q}
    \big\|
        \mathsf{B}_\theta\,\bm{c}^{(N_q)}_\mu(\theta) - \bm{f}
    \big\|_2^2.
\end{equation}

\subsection{Discretization of the Green integral operator $\mathcal{G}$}
\label{sec:green_discretization}
Since $\mathcal{B} = \mathcal{I} - \mathcal{G}$, the main computational task is the
evaluation of the Green integral operator $\mathcal{G}$ defined in~\eqref{eq:green_operator}.
We discretize $\mathcal{G}$ on the same uniform grid $\{x_k\}_{k=1}^{N_q} \subset \Omega$
introduced above, with grid spacings $\Delta x$ and $\Delta z$. For clarity, the formulas
below are written for $d = 2$; the three-dimensional case is analogous, with the cell
volume $\Delta x\,\Delta z$ replaced by $\Delta x\,\Delta y\,\Delta z$. For each basis
function $\varphi_j$, the action of $\mathcal{G}$ at a grid point $x_k$ is approximated by
the composite midpoint rule
\begin{equation}
\label{eq:green_quadrature}
    \mathcal{G}\varphi_j(x_k;\theta)
    \;\approx\;
    \sum_{\ell=1}^{N_q}
    g_0(x_k - x_\ell)\,
    \underbrace{
      \Delta x\,\Delta z\;\omega^2\,
      \delta m(x_\ell)\,
      \varphi_j(x_\ell;\theta)
    }_{(F_\theta)_{\ell j}},
    \qquad k = 1, \ldots, N_q.
\end{equation}
The underbraced term collects the quadrature weight, the scattering potential, and the
basis function into a single weighted-basis matrix
\begin{equation}
\label{eq:weighted_basis}
    F_\theta
    \;=\;
    \Delta x\,\Delta z\;\omega^2\,
    \operatorname{diag}(\delta\bm{m})\,\Phi_\theta
    \;\in\; \mathbb{C}^{N_q \times N},
\end{equation}
where $\delta\bm{m} = (\delta m(x_1), \ldots, \delta m(x_{N_q}))^\top$ and
$\Phi_\theta \in \mathbb{C}^{N_q \times N}$ has entries $(\Phi_\theta)_{kj} = \varphi_j(x_k;\theta)$.
With this notation, the quadrature rule~\eqref{eq:green_quadrature} for all $j$ and $k$
takes the compact matrix form
\begin{equation}
\label{eq:green_matrix}
    \mathcal{G}\Phi_\theta
    \;=\;
    \bm{g}_0\,F_\theta
    \;\in\; \mathbb{C}^{N_q \times N},
\end{equation}
where $\bm{g}_0 \in \mathbb{C}^{N_q \times N_q}$ is the discrete Green matrix with entries
$(\bm{g}_0)_{k\ell} = g_0(x_k - x_\ell)$. Because the grid is uniform, $\bm{g}_0$ is a
Toeplitz matrix, and~\eqref{eq:green_matrix} applies it to each column of $F_\theta$
independently:
\begin{equation}
\label{eq:green_column}
(\mathcal{G}\Phi_\theta)_{\cdot,j}
    \;=\;
    \bm{g}_0\,(F_\theta)_{\cdot,j},
    \qquad j = 1, \ldots, N.
\end{equation}
A Toeplitz matrix-vector product is precisely a discrete linear convolution. Thus  
each column of~\eqref{eq:green_matrix} can equivalently be evaluated by convolving
$g_0$ with the corresponding column of $F_\theta$; applied column by column. This gives
\begin{equation}
\label{eq:green_convolution}
    \mathcal{G}\Phi_\theta
    \;=\;
    \bm{g}_0 * F_\theta.
\end{equation}
This convolution structure is what enables an FFT-accelerated evaluation of $\mathcal{G}\Phi_\theta$,
avoiding the dense $O(N_q^2)$ cost of forming and applying $\bm{g}_0$ directly. Once
$\mathcal{G}\Phi_\theta$ is computed, the discrete operator $\mathsf{B}_\theta$ is recovered as
\begin{equation}
\label{eq:discrete_B}
    \mathsf{B}_\theta = \Phi_\theta - \mathcal{G}\Phi_\theta
    \;\in\; \mathbb{C}^{N_q \times N}.
\end{equation}

\begin{remark}
The Green's function kernel $g_0$ is singular at the origin, but the quadrature formula
\eqref{eq:green_quadrature} requires its value there. We
regularize this singular value using a circular-cell strategy or a rectangular-cell
approximation. Both give a similar accuracy, so we adopt the circular-cell regularization
throughout. See  Appendix~\ref{appendix:Regularization of the free-space Green's function singularity} for details.
\end{remark}

\subsection{Adaptive-depth strategy}
\label{subsec:adaptive_depth}

Scattered wave propagation is a highly multiscale problem: spectral bias hinders efficient training of deep networks that must capture the fine scales of the wavefield. Training a deep network of the full target depth $L$ from random initialization can be challenging in strong scattering regimes due to the highly non-convex, oscillatory nature of the loss landscape. We instead develop an adaptive-depth training strategy for the plane-wave basis. We train a sequence of sub-networks of increasing depth $\ell = 1, 2, \ldots, L$, each initialized from the converged parameters of the previous stage. This preserves the first- and second-order moments of Adam, and the linear coefficients of all neurons forming the basis are recovered exactly at every step, which together avoid spikes in the loss during training. We explain the algorithm as follows.

\begin{figure}[!h]
    \centering
    \includegraphics[width=1.0\linewidth]{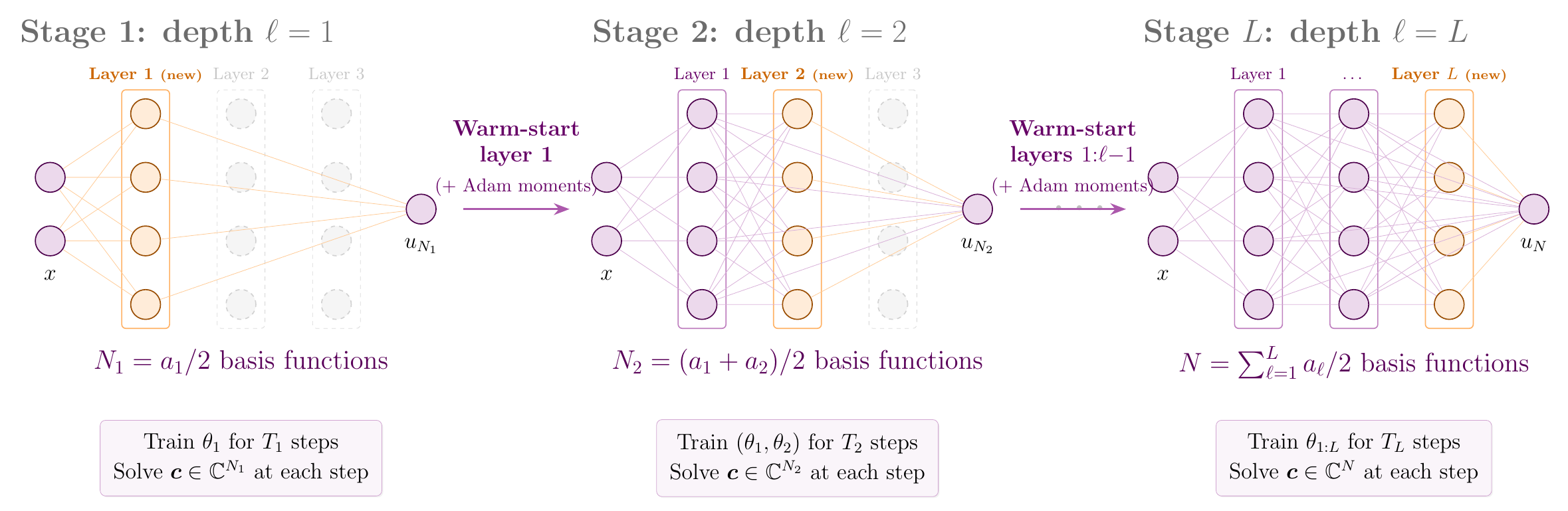}
    \caption{Adaptive-depth strategy. At each stage $\ell$, only the first $\ell$ hidden layers are active (violet) and the newly introduced layer (orange) is randomly initialized, while layers $1, \ldots, \ell{-}1$ are warm-started from the previous stage. All basis functions from all active layers are connected to the output, and the linear coefficients $\bm{c} \in \mathbb{C}^{N_\ell}$ are recomputed by solving the normal equations at each step. The number of basis functions grows as $N_\ell = \sum_{i=1}^\ell a_i/2$.}
    \label{fig:adaptive_depth}
\end{figure}

We start training with a network of a single layer, giving a coarse first basis, and grow the network one layer at a time. At depth $\ell$, only the first $\ell$ layers are active and contribute basis functions, so the active basis grows as $N_\ell = \sum_{i=1}^\ell a_i/2$. At each depth $\ell$ we run a fixed number of optimizer steps $T_\ell$, solving the normal equations~\eqref{eq:discrete_normal_equations} at every step. When depth increases from $\ell$ to $\ell+1$, the parameters of the first $\ell$ layers, together with their first- and second-order Adam moments, are carried over unchanged; only the new layer is randomly initialized. Resetting the moments would make Adam treat the converged parameters as freshly initialized, discarding their accumulated curvature information and causing a sharp increase in the loss. This transfer of both parameters and optimizer state acts as a form of transfer learning across depths, and it is what provides a good initialization at every successive depth as the network grows. A schematic overview is given in Figure~\ref{fig:adaptive_depth}.

Adaptive-depth training has been explored previously in the data-driven setting~\cite{wen2020autogrow, krishnanunni2025topological}. We develop it here for a different setting, with two consequences for how it must be designed. First, each added layer contributes a new set of neural basis functions rather than standard weights: the corresponding linear coefficients are recovered by solving a least-squares problem at every step, not learned directly, so growing the network grows the basis itself. Second, the target is a scientific computing problem governed by a physics-based loss, not a data-driven task, so no ground-truth field is ever seen during training.

This second point drives the central design question of any depth-growing scheme: when to add a new layer, and how long to train at each depth. In a data-driven setting this decision can be guided directly, using a validation set to trigger growth once the model stops improving against the reference solution. In our setting no such reference exists: the only supervisory signal is the GI loss, which is physics-informed and never sees the ground-truth field. Since the decision cannot be made adaptively from a validation signal, we design it in advance instead: we fix the target depth $L$ and a training schedule $(T_1, \ldots, T_L)$ ahead of training, with $\sum_{\ell=1}^{L} T_\ell = T_{\mathrm{total}}$ the total iteration budget.

We allocate this budget according to what we call the \textit{heavy-tail} schedule: the iteration count increases with depth, $T_1 \le T_2 \le \cdots \le T_L$. Layer $1$, the cheapest sub-network to optimize, is allotted the fewest iterations; deeper, harder-to-optimize sub-networks receive proportionally more iterations to converge before the next layer is added. The network parameters $\theta$ are updated with a learning rate initialized at $10^{-3}$ and decayed exponentially, with gradient norm clipping applied at a threshold of one.

\section{Algorithmic and computational aspects}
\label{sec:implementation}

\begin{figure}[!h]
    \centering
\includegraphics[width=1.0\linewidth]{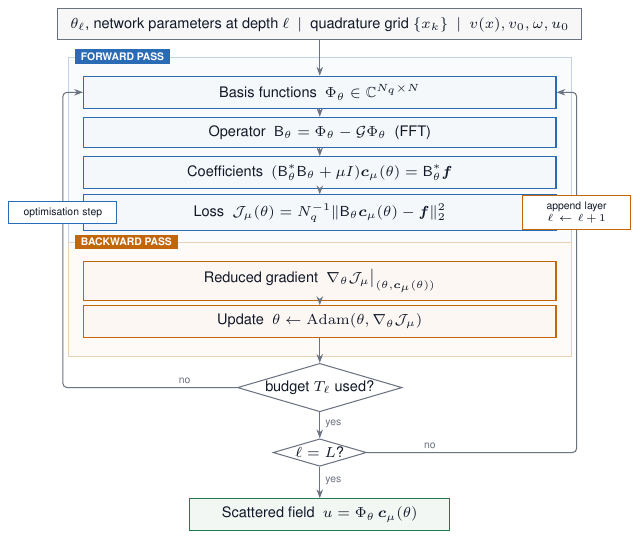}
    \caption{General workflow of the NAP-Wave solver. Each forward pass evaluates
the loss $\mathcal{J}_\mu(\theta)$: for fixed $\theta$, the basis functions
$\varphi_j$ are evaluated on the quadrature grid, $\mathsf{B}_\theta$ is
assembled via the FFT-accelerated Green operator, and $\bm{c}$ is obtained by
solving the Tikhonov-regularized normal equations through a Cholesky
factorization. In the backward pass, the reduced gradient
$\nabla_\theta \mathcal{J}_\mu$ is computed without differentiating through
$\bm{c}_\mu(\theta)$ and used to update $\theta$ with Adam. The inner loop
repeats this for the $T_\ell$ steps allotted to the current depth $\ell$; the
outer loop then appends a layer and warm-starts the next stage, until the
target depth $L$ is reached.}
    \label{fig:general_workflow}
\end{figure}

In the following, we describe the main implementation aspects of the NAP-Wave
solver. The full procedure is summarized in Algorithm~\ref{alg:hybrid_optimization},
and the general workflow is illustrated in Figure~\ref{fig:general_workflow}. We
focus on the following aspects: the efficient assembly of the discrete normal-equation
matrices in  \eqref{eq:discrete_normal_equations}, the linear
solver, the choice of backend library, and numerical precision.

\subsection{Computational domain}
\label{subsec:computational_domain}

\begin{figure}[ht]
    \centering
    \includegraphics[width=1.0\linewidth]{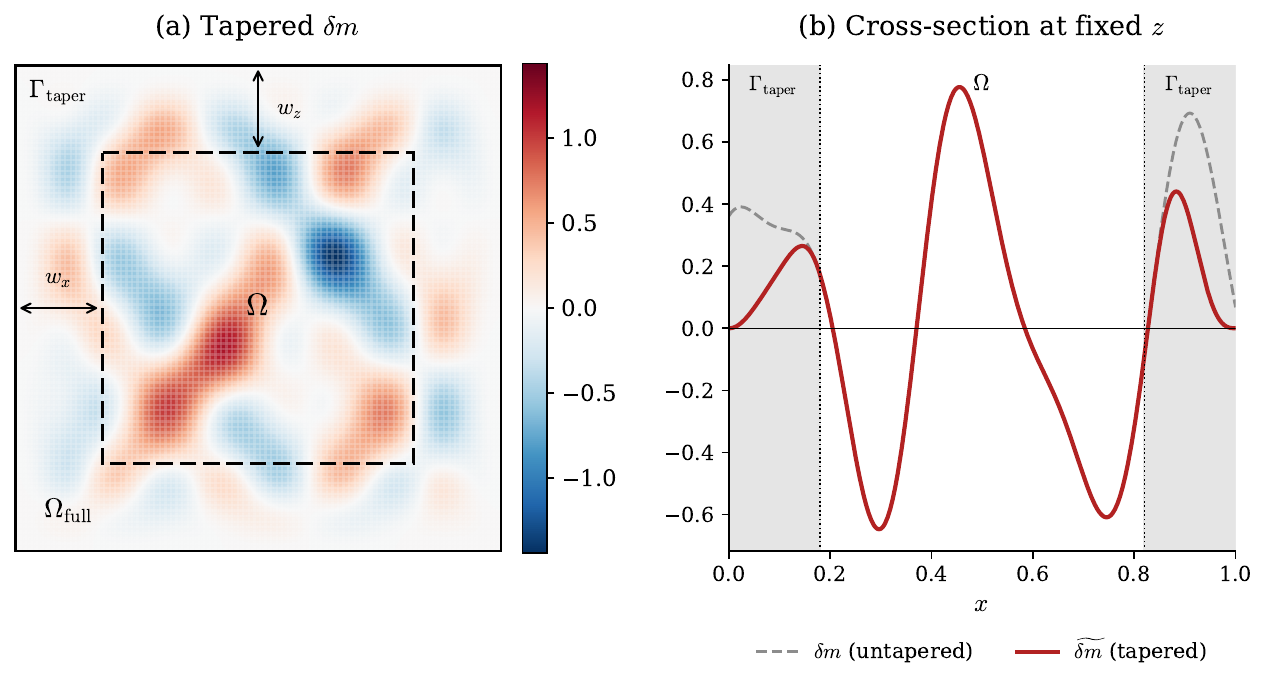}
    \caption{Construction of the full computational domain $\Omega_{\mathrm{full}}$ and the tapering
    of the medium contrast. (a) The tapered contrast $\widetilde{\delta m}$ over
    $\Omega_{\mathrm{full}}$: the original domain $\Omega$ is embedded in a larger domain
    surrounded by a border region $\Gamma_{\mathrm{taper}}$ of width $w_x$ in the horizontal
    direction and $w_z$ in the vertical direction. (b) Cross-section at fixed $z$ comparing the
    untapered contrast $\delta m$ with $\widetilde{\delta m}$.}
    \label{fig:tapering_delta_m}
\end{figure}

One of the main practical challenges is that we only have access to the velocity
field on a bounded domain $\Omega$, whereas the theoretical formulation of
Section~\ref{sec:mathematical_models} requires the contrast $\delta m$ to have
compact support strictly inside $\Omega$. To resolve this, we embed $\Omega$
inside a larger computational domain
$\Omega_{\mathrm{full}} = \Omega \cup \Gamma_w$,
where $\Gamma_w$ denotes a padding (taper) region. Within $\Gamma_w$, the contrast
is smoothly damped to zero using a raised-cosine taper (see
Appendix~\ref{appendix:compact_support}), such that the tapered contrast
$\widetilde{\delta m}$ satisfies
$\operatorname{supp}(\widetilde{\delta m}) \subset \Omega_{\mathrm{full}}$. A
schematic overview is given in Figure~\ref{fig:tapering_delta_m}.

We adopt a nondimensionalization form of the
Lippmann-Schwinger integral equation. The physical coordinates $x$ and $z$, the
velocities $v$ and $v_0$, are rescaled by $\lambda_0 = 2\pi v_0/\omega$, giving
$v_{\mathrm{nd}} = v/\lambda_0$ and $v_{0,\mathrm{nd}} = v_0/\lambda_0$. The nondimensional
contrast then becomes
\begin{equation}
    \delta m_{\mathrm{nd}}(x)
    = \frac{1}{v_{\mathrm{nd}}(x)^2} - \frac{1}{v_{0,\mathrm{nd}}^2}
    = \lambda_0^2\left(\frac{1}{v(x)^2} - \frac{1}{v_0^2}\right)
    = \lambda_0^2 \;\delta m(x).
\end{equation}

\subsection{Network architecture}
\label{subsec:arch_FNN}
We use a sine-activated feedforward network whose hidden activations are
reorganized into complex basis functions, as detailed in
Appendix~\ref{appendix:arch} and shown in Figure~\ref{fig:basis_construction}.
Sine activations provide  infinitely differentiable features that are well
suited to representing oscillatory wave fields. We initialize the neural
parameters $\theta = \{\, W^{(i)}, \bm{b}^{(i)} \,\}_{i=1}^L$ with a uniform
scheme: each weight matrix is drawn entrywise i.i.d.\ from
\begin{equation}
    W^{(i)}_{kl}
    \;\sim\;
    \mathcal{U}\!\left(
        -\sqrt{\tfrac{6}{d_{i-1}}},\;
        \sqrt{\tfrac{6}{d_{i-1}}}
    \right),
    \qquad i = 1,\ldots,L,
    \label{eq:init_W}
\end{equation}
and the biases are initialized to zero, $\bm{b}^{(i)} = \bm{0} \in \RR^{a_i}$.

The plane-wave basis~\eqref{eq:plane_wave_basis} is built from complex
carriers $\chi_i(x) = \exp\!\bigl( \mathrm{i}\, k_i\, \bm{d}_i \cdot x \bigr)$,
each defined by a unit direction vector $\bm{d}_i \in \mathbb{S}^{d-1}$ sampled
deterministically and uniformly over the sphere
\begin{equation}
    \bm{d}_i
    \;=\;
    \begin{cases}
        \bigl(\cos \alpha_i,\; \sin \alpha_i\bigr),
        & d = 2, \\[2mm]
        \bigl(\rho_i \cos \beta_i,\; \rho_i \sin \beta_i,\; \zeta_i\bigr),
        & d = 3,
    \end{cases}
    \qquad i = 1,\dots,N,
    \label{eq:pw_directions}
\end{equation}
where, for $d = 2$, $\alpha_i = 2\pi (i-1)/N$, and for $d = 3$,
\begin{equation}
   \displaystyle \zeta_i = 1 - \frac{2i-1}{N},
    \quad
    \rho_i = \sqrt{1 - \zeta_i^{\,2}},
    \quad
    \beta_i = 2\pi (i-1)\,\left(\tfrac{1+\sqrt{5}}{2}\right)^{-2},
\label{eq:pw_directions_3d}
\end{equation}
and a positive wavenumber
\begin{equation}
    k_i
    \;=\; \mathrm{softplus}\bigl( \tilde k_i \bigr) + k_{\min},
    \qquad k_{\min} > 0,
    \label{eq:pw_wavenumber}
\end{equation}
where $\tilde k_i \in \RR$ is trainable. We adopt this
softplus-plus-$k_{\min}$ parametrization to guarantee $k_i > k_{\min} > 0$
for any $\tilde k_i \in \RR$, so the wavenumber remains strictly positive and
bounded away from zero throughout training. We initialize each
$\tilde k_i$ from the true local wavenumbers $k(x)$ over the velocity
grid, restricted to the quantile range $[q_{0.05}, q_{0.95}]$ to exclude
outliers.

\subsection{Efficient solving of the normal equations}
\label{subsec:fft_assembly}
The dominant cost at each iteration of the NAP-Wave solver is the construction of the matrix
$\mathcal{G}\Phi_{\theta_n}$ given by~\eqref{eq:green_convolution}. To implement
this efficiently, we exploit the convolution theorem via the fast Fourier
transform, which yields
\begin{equation}
\label{eq:green_fft}
    \mathcal{G}\Phi_\theta
    =
    \mathcal{F}^{-1}
    \!\left(
        \mathcal{F}[\bm{g}_0]
        \odot
        \mathcal{F}[F_\theta]
    \right),
\end{equation}
where $\mathcal{F}$ denotes the discrete Fourier transform and $\mathcal{F}^{-1}$
its inverse. This reduces the cost of computing~\eqref{eq:green_convolution} from
$O(N N_q^2)$ to $O(N N_q \log N_q)$ per iteration. Implementation details are
given in Appendix~\ref{appendix:fft_acceleration}. The transform
$\mathcal{F}[\bm{g}_0]$ is computed once, at initialization.

For the regularization parameter $\mu_n > 0$, we adopt an exponential decay
schedule:
\begin{equation}
    \mu_n \;=\; \mu_{0}
                 \left(\frac{\mu_{f}}{\mu_{0}}\right)^{ n / N_{f}},
\end{equation}
where $n$ the current iteration
and $N_{f}$ the total number of iterations, where with
$\mu_{0} = 10^{-3}$ and $\mu_{f} = 10^{-6}$.

After assembling $\mathsf{B}_{\theta_n}^* \mathsf{B}_{\theta_n} + \mu_n I$, the
system is solved via Cholesky factorization followed by two triangular solves. A direct solver is preferred because the
system is solved exactly once per iteration, so the $O(N^3)$ factorization cost
is amortized against the gradient computation.

The linear system~\eqref{eq:discrete_normal_equations} is solved directly using complex-valued
arithmetic, rather than converting it into an equivalent real-valued block
system, which incurs additional computational cost as was implemented in \cite{abedi2026least}.
We note that we also provide a real-valued mode for devices or libraries that
do not offer native support for complex-valued operations. We found that
solving directly in complex-valued arithmetic is faster than using the
equivalent real-valued block system, by a factor of approximately $2.7$.

\subsection{Backend library and float precision}
\label{subsec:backend}

The implementation is built on \texttt{JAX}, selected over \texttt{PyTorch} and
\texttt{TensorFlow} after extensive preliminary benchmarks. Two features of
\texttt{JAX} are particularly relevant to our algorithm. First, the inner training loop consists of a fixed sequence of dense
linear-algebra operations (basis evaluation, FFT, Cholesky factorization,
triangular solves, matrix-vector products). Wrapping each iteration in
\texttt{jax.jit} fuses these operations into a single optimized XLA
computational graph, removing Python-level overhead. Second, the neural basis $\{\varphi_j(\cdot\,;\theta)\}_{j=1}^N$ is expressed as a single network evaluated over $N$ output channels and $N_q$
quadrature nodes. We use \texttt{jax.vmap} to vectorize both axes without
explicit batched code, keeping the implementation close to the mathematical
formulation while delivering the throughput of a hand-written batched kernel. 

All computations are carried out in double precision (\texttt{float64}).
Double precision is required for the Cholesky factorization where a small $\mu$ or
nearly linearly dependent basis functions render the system matrix
ill-conditioned, causing the factorization to lose positive definiteness in
single precision.

\begin{algorithm}[!h]
\caption{NAP-Wave solver with adaptive-depth training strategy}
\label{alg:hybrid_optimization}
\begin{algorithmic}[1]
\Require Initial parameters $\theta^0 \in \RR^p$, target network depth $L$,
iterations per depth $\{T_\ell\}_{\ell=1}^L$, regularization schedule
$\mu_n > 0$, quadrature points $(x_i)_{i=1}^{N_q}$, source vector $\bm{f} \in \mathbb{C}^{N_q}$
\Ensure Approximate scattered field $u_N \in \mathcal{V}_N$
\State Precompute and store $\mathcal{F}[\bm{g}_0]$ on the zero-padded grid
\State Compute the contrast vector $\delta \bm m$ on the quadrature points
\For{$\ell = 1, \ldots, L$} \Comment{Adaptive-depth loop}
    \State Set $N_\ell = \sum_{i=1}^\ell a_i/2$ basis functions
    \If{$\ell > 1$}
        \State Warm-start layers $1, \ldots, \ell-1$ from the previous stage, along with their first- and second-order Adam moments
        \State Initialize layer $\ell$ randomly
    \EndIf
    \For{$n = 0, \ldots, T_\ell - 1$}
        \State Evaluate $\Phi_{\theta_n} \in \mathbb{C}^{N_q \times N_\ell}$ on the grid
        \State Assemble $\mathsf{B}_{\theta_n} = \Phi_{\theta_n} -
        \mathcal{G}\Phi_{\theta_n}$ via FFT~\eqref{eq:green_fft}
        \State Factorize
        $\mathsf{B}_{\theta_n}^* \mathsf{B}_{\theta_n} + \mu_n I = CC^*$
        (Cholesky)
        \State Solve for $\bm{c}^{(N_q)}_{\mu_n}(\theta_n)$
        \State Evaluate $\mathcal{J}^{(N_q)}_{\mu_n}(\theta_n) =
        \frac{1}{N_q}
        \big\|\mathsf{B}_{\theta_n}\,\bm{c}^{(N_q)}_{\mu_n}(\theta_n)
        - \bm{f}\big\|_2^2$
        \State Compute $\nabla_\theta \mathcal{J}^{(N_q)}_{\mu_n}(\theta_n)$
        \State Update: $\theta_{n+1} = \mathrm{Adam}\big(\theta_n,\,
        \nabla_\theta \mathcal{J}^{(N_q)}_{\mu_n}(\theta_n)\big)$
    \EndFor
\EndFor
\State \Return $u_N(x) = \displaystyle\sum_{j=1}^{N}
\big(\bm{c}^{(N_q)}_\mu(\theta^{\mathrm{final}})\big)_j\,
\varphi_j(x;\theta^{\mathrm{final}})$
\end{algorithmic}
\end{algorithm}

\section{Case studies}
\label{sec:case_studies}

In this section, we present numerical experiments to assess the accuracy and computational efficiency of the proposed NAP-Wave solver. We consider five benchmarks. The first two, scattering by a fluid cylinder in 2D and a linear radial velocity gradient in 3D, admit closed-form analytical solutions and serve to verify the method against exact references. 

\begin{figure}[!h]
    \centering
    \includegraphics[width=1.0\linewidth]{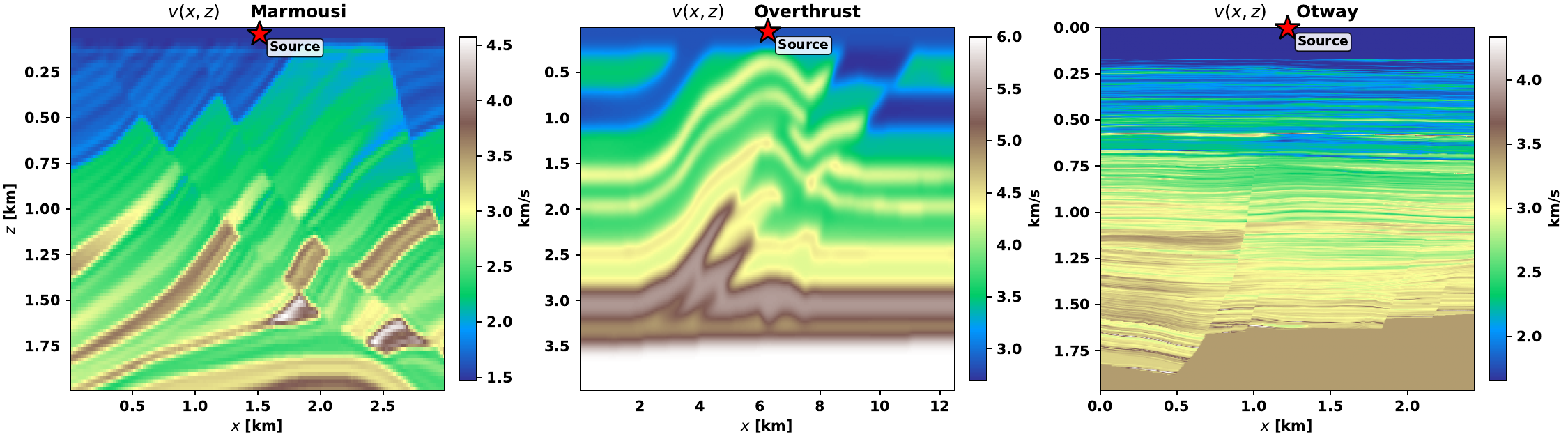}
    \caption{Velocity fields $v(x,z)$ of the three benchmark models: Marmousi (left), Overthrust (middle), and Otway (right). The corresponding parameters are summarized in Table~\ref{tab:numerical_test_summary}.}
    \label{fig:velocity_profiles}
\end{figure}

The remaining three benchmarks, Marmousi, Overthrust, and Otway, are highly heterogeneous, multiscale velocity models originally derived from real geophysical structural data. These datasets are standard in the seismic imaging community, and we adopt them here as rigorous tests for neural PDE solvers. They exhibit sharp, non-smooth property contrasts, complex stratigraphic layering, and strong multiscale heterogeneities. These features cause severe spectral bias in standard neural networks and create poorly conditioned loss landscapes, making them difficult benchmark datasets for machine learning solvers to fit accurately without specialized architectures. The velocity field and source location for each of these three benchmarks are illustrated in Figure~\ref{fig:velocity_profiles}; a summary of the numerical parameters is given in Table~\ref{tab:numerical_test_summary}, and detailed descriptions are provided in the following subsections. 

\begin{table}[!h]
    \centering
    \scriptsize
    \caption{Summary of the numerical benchmarks considered in this paper.}
    \label{tab:numerical_test_summary}
    \vspace{4pt}
    \scriptsize
    \tablestyle
    \setlength{\tabcolsep}{1pt}
    \begin{tabular}{|l|c|c|c|c|c|}
        \hline
        \rowcolor{tblheaderblue}
        \textbf{Parameter} & \textbf{Fluid-cylinder scattering} & \textbf{3D radial velocity gradient} & \textbf{Marmousi} & \textbf{Overthrust} & \textbf{Otway} \\
        \hline
        Domain $\Omega$ (km)
            & $[0,3.0]\times[0,2.0]$
            & $[-1,1]^3$
            & $[0,3.0]\times[0,2.0]$
            & $[0,12.5]\times[0,4.0]$
            & $[0,2.438]\times[0,1.966]$ \\
        \hline
        \rowcolor{tblrowblue}
        Background velocity $v_0$ (km/s) & $1.5$ & $1.5$ & $1.5$ & $2.856$ & $1.65$ \\
        \hline
        Source position $x_{\mathrm{s}}$ (km) & $(0.3, 0.1)$ & $(0,0,0)$ & $(1.513, 0.042)$ & $(6.265, 0.052)$ & $(1.2204, 0.0036)$ \\
        \hline
        \rowcolor{tblrowblue}
        Frequency $f$ (Hz) & $6$ & $2$ & $10$ & $10$ & $20$ \\
        \hline
        Background wavenumber $k_0$ (km$^{-1}$) & $\approx 25.13$ & $\approx 8.38$ & $\approx 41.89$ & $\approx 22.00$ & $\approx 76.16$ \\
        \hline
    \end{tabular}
\end{table}

\subsection{Exact solution benchmarks}

We consider two benchmark problems with closed-form exact solutions.

\paragraph{2D fluid-cylinder scattering.}
We consider a disk of radius $a = 0.3$ centered at $(1.5, 1.0)$, with
constant velocity $v_1 = 2.25$, embedded in a homogeneous background
$v_0 = 1.5$ over the domain $[0,3]\times[0,2]$, excited at frequency
$f = 6\,\mathrm{Hz}$ by the line-source incident field $u_0$ located at
$x_s = (0.3, 0.1)$. The exact solution, expressed in terms of the total field reads
\begin{equation}
    u(r,\phi) =
    \begin{cases}
        \displaystyle\sum_{n\in \mathbb{Z}} c_n \left[J_n(k_0 r) + a_n H_n^{(2)}(k_0 r)\right]e^{in\phi}, & r > a, \\[2mm]
        \displaystyle\sum_{n\in \mathbb{Z}} c_n\, b_n\, J_n(k_1 r)\, e^{in\phi}, & r < a,
    \end{cases}
\end{equation}
where $(r,\phi)$ are polar coordinates centered at the disk, $k_0=\omega/v_0$, $k_1=\omega/v_1$, and $c_n$ is the incident field's local multipole amplitude at the disk. The coefficients $a_n,b_n$ follow from matching $u$ and its normal derivative at $r=a$.

\paragraph{3D radial velocity gradient.}
The velocity increases linearly with distance from the source $x_s = (0,0,0)$,
\begin{equation}
    v(r) = v_0 + \alpha r, \qquad r = \|x - x_s\|,\ \alpha > 0,
\end{equation}
with $v_0 = 1.5$ and $\alpha = 0.6$, over the domain $[-1,1]^3$ at
frequency $f = 2\,\mathrm{Hz}$. The exact solution, expressed in terms of the total field reads
\begin{equation}
    u(r) = \frac{1}{4\pi r}\sqrt{\frac{v(r)}{v_0}}\left(\frac{v_0}{v(r)}\right)^{i\tilde{\mu}}, \qquad \tilde{\mu} = \sqrt{\left(\frac{\omega}{\alpha}\right)^2 - \frac{1}{4}},
\end{equation}

\subsection{Marmousi benchmark} 
We consider the Marmousi model, a synthetic acoustic velocity model used as a
benchmark for seismic imaging and full-waveform inversion in heterogeneous
media \cite{brougois1990marmousi}. The geometric and physical parameters of this benchmark are
summarized in Table~\ref{tab:numerical_test_summary}, and the
corresponding velocity field $v(x, z)$ is shown in
Figure~\ref{fig:velocity_profiles}. The model is representative of complex
sedimentary structures, with a strongly heterogeneous, multi-layered
velocity field exhibiting stratigraphic complexity, dipping reflectors, and
sharp contrasts across geological interfaces. These features make it a
challenging configuration for wave-scattering solvers, as the resulting
wavefield combines high-frequency oscillations with localized reflections
and refractions.

\subsection{Overthrust benchmark} 
We consider the Overthrust model~\cite{aminzadeh1994seg}, a synthetic acoustic
benchmark released by the Society of Exploration Geophysicists (SEG) and the
European Association of Geoscientists and Engineers (EAGE), used as a
reference for seismic imaging in structurally complex, faulted geological
settings. The
geometric and physical parameters of this benchmark are summarized in
Table~\ref{tab:numerical_test_summary}, and the corresponding velocity
field $v(x, z)$ is shown in Figure~\ref{fig:velocity_profiles}. The model
represents a thrust-belt geological setting, characterized by a layered
sedimentary sequence disrupted by overthrust faulting, producing offset
velocity discontinuities and locally steep, non-conformable interfaces.
Compared to the Marmousi model, the Overthrust velocity field exhibits
smoother lateral variations within each layer but sharper,
structurally-driven discontinuities across fault planes, making it a
complementary test for assessing the sensitivity of wave-scattering
solvers to abrupt structural offsets rather than fine-scale stratigraphic
heterogeneity.

\subsection{Otway benchmark} As a last benchmark, we consider the
Otway model, a synthetic acoustic velocity model built from geological and
well-log data of the Otway Basin, Australia, a sedimentary basin widely
studied in the context of geological carbon storage and CO$_2$
sequestration monitoring. The geometric and physical parameters of this
benchmark are summarized in Table~\ref{tab:numerical_test_summary}, and the
corresponding velocity field $v(x, z)$ is shown in
Figure~\ref{fig:velocity_profiles}. The model features a comparatively
smooth, gently layered sedimentary sequence with moderate velocity
contrasts between formations, in contrast with the strong heterogeneity and
stratigraphic complexity of the Marmousi and Overthrust models. This
benchmark is therefore representative of near-surface, moderately
heterogeneous sedimentary basins, and allows us to assess the accuracy and
efficiency of the proposed solver in a regime dominated by smoother,
large-scale velocity variations rather than sharp interfaces.

\subsection{Reference solutions, reproducibility, and hardware}

For velocity models admitting a closed-form solution, we compare directly
against the exact solution. For the realistic velocity models, we generate a
numerical reference using an optimal nine-point finite-difference discretization of the
scattered-field Helmholtz equation with a perfectly matched layer (PML). We compute the relative $L^2$ error on a uniform grid as
\begin{equation}
    \mathrm{Relative}\;L^2
    \;(\text{in }\%)\;=\;
    100\times \left(\frac{\sum_{i} \bigl|\, u_{\mathrm{pred}}(x_i) - u_{\mathrm{ref}}(x_i) \,\bigr|^2}
         {\sum_{i} \bigl|\, u_{\mathrm{ref}}(x_i) \,\bigr|^2}\right)^{1/2},
    \label{eq:rel_l2}
\end{equation}
computed over the interior (taper-excluded) region.

All experiments are performed on a single NVIDIA H100 80GB HBM3 GPU. The implementation uses \texttt{JAX}~0.9.1 with \texttt{Flax}~0.12.6 and \texttt{Optax}~0.2.4 for automatic differentiation and optimization, and \texttt{NumPy}~2.1.1 and \texttt{SciPy}~1.17.0 for numerical routines, running on Python~3.11.5 with CUDA~12 support. The source code is publicly available at \url{https://github.com/Hamza-Kamil/NAP-WAVE}.

\section{Numerical results}
\label{sec:numerical_results}
\subsection{Qualitative validation}
\label{sec:qualitative_validation}

For each benchmark, we depict the real and imaginary parts of the predicted and reference scattered field, together with the pointwise error. All fields are displayed in the nondimensional coordinates. The network architecture and optimizer settings are summarized in Table~\ref{tab:configuration_network_training}.

\begin{table}[!h]
    \centering
     \small
\caption{Network architecture, training, and evaluation settings used to produce
Figures~\ref{fig:Exact_2D_predictions}-\ref{fig:Otway_predictions_full} and
Table~\ref{tab:comparison_literature}. The quadrature grid and test grid are given as
$N_x \times N_z$ in 2D and $N_x \times N_y \times N_z$ in 3D, where $N_x$, $N_y$, and
$N_z$ denote the number of points along the $x$-, $y$-, and $z$-directions, respectively.
The total number of iterations is the sum of the iterations allocated to each depth.}
    \label{tab:configuration_network_training}
    \vspace{4pt}
    \tablestyle
    \setlength{\tabcolsep}{6pt}
    \begin{tabular}{|l|c|c|c|c|c|}
        \hline
        \rowcolor{tblheaderblue}
        \textbf{Setting} & \textbf{Fluid-cylinder scattering} & \textbf{3D radial velocity gradient} & \textbf{Marmousi} & \textbf{Overthrust} & \textbf{Otway} \\
        \hline
        Depth / width & $5/128$ & $3/128$ & $5/128$ & $5/128$ & $5/128$ \\
        \hline \rowcolor{tblrowblue}
        Basis functions $N$ & $320$ & $192$ & $320$ & $320$ & $320$ \\
        \hline
        Quadrature grid $N_q$ & $340\!\times\!240$ & $60\times60\times60$ & $180\!\times\!260$ & $512\!\times\!160$ & $256\!\times\!200$ \\
        \hline \rowcolor{tblrowblue}
        \hline
        Total iterations & $20{,}000$ & $25{,}100$ & $26{,}100$ & $100{,}001$ & $200{,}001$ \\
        \hline 
        Test grid & $150\!\times\!100$ & $20\times 20 \times 20$ & $150\!\times\!100$ & $500\!\times\!160$ & $ 610\!\times\!492$ \\
        \hline
    \end{tabular}
\end{table}

\subsubsection{Exact solutions}

\begin{figure}[!h]
    \centering
    \includegraphics[width=1.0\linewidth]{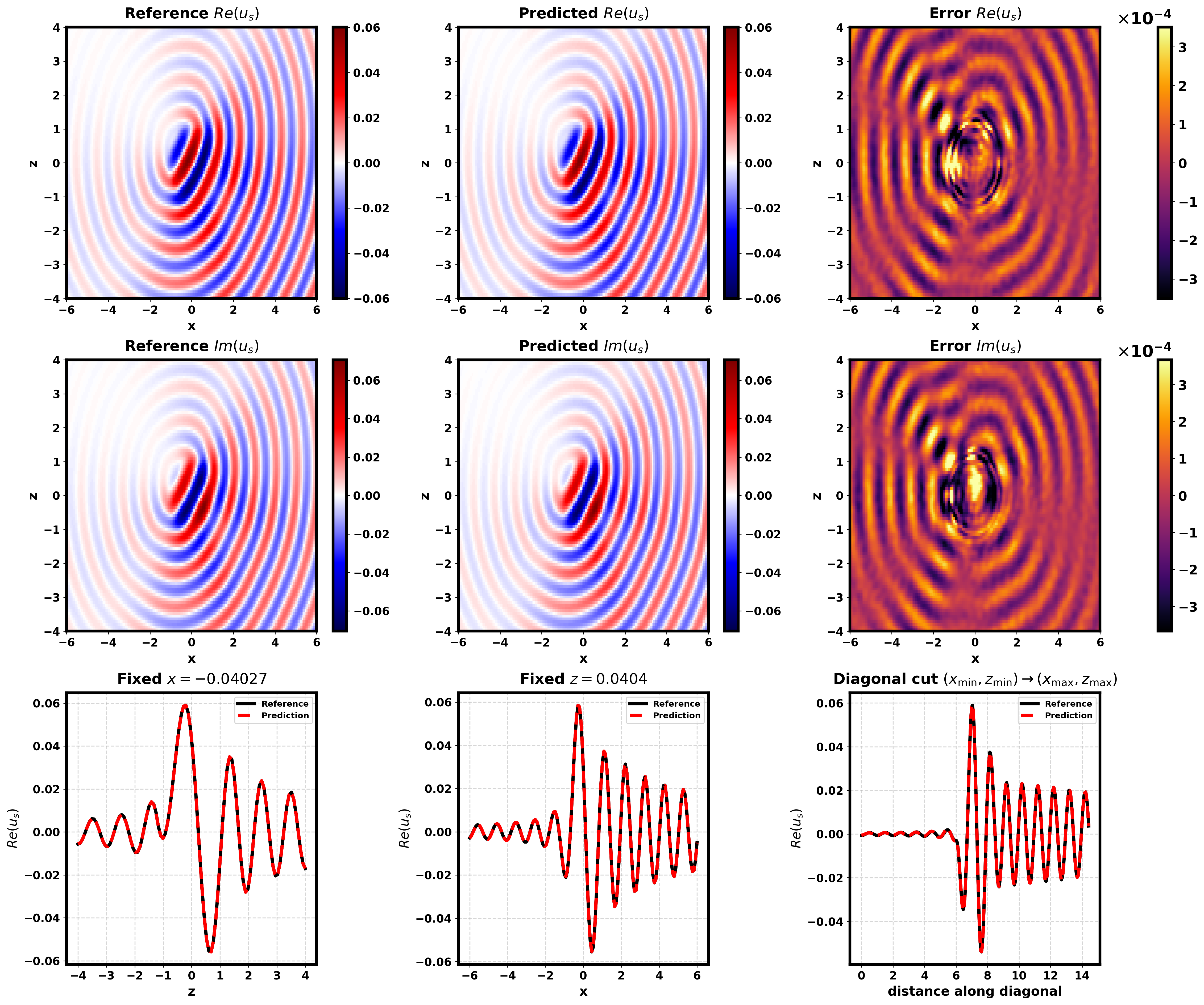}
    \caption{Fluid-cylinder scattering  benchmark: real part (top row) and imaginary part (middle row) of the scattered field $u_s$, comparing the reference, the prediction, and the pointwise error. Bottom row: cross-sections along fixed $x$, fixed $z$, and the domain diagonal. Relative $L^2$ error: $0.97\%$.}
    \label{fig:Exact_2D_predictions}
\end{figure}

\begin{figure}[!h]
    \centering
\includegraphics[width=1.0\linewidth]{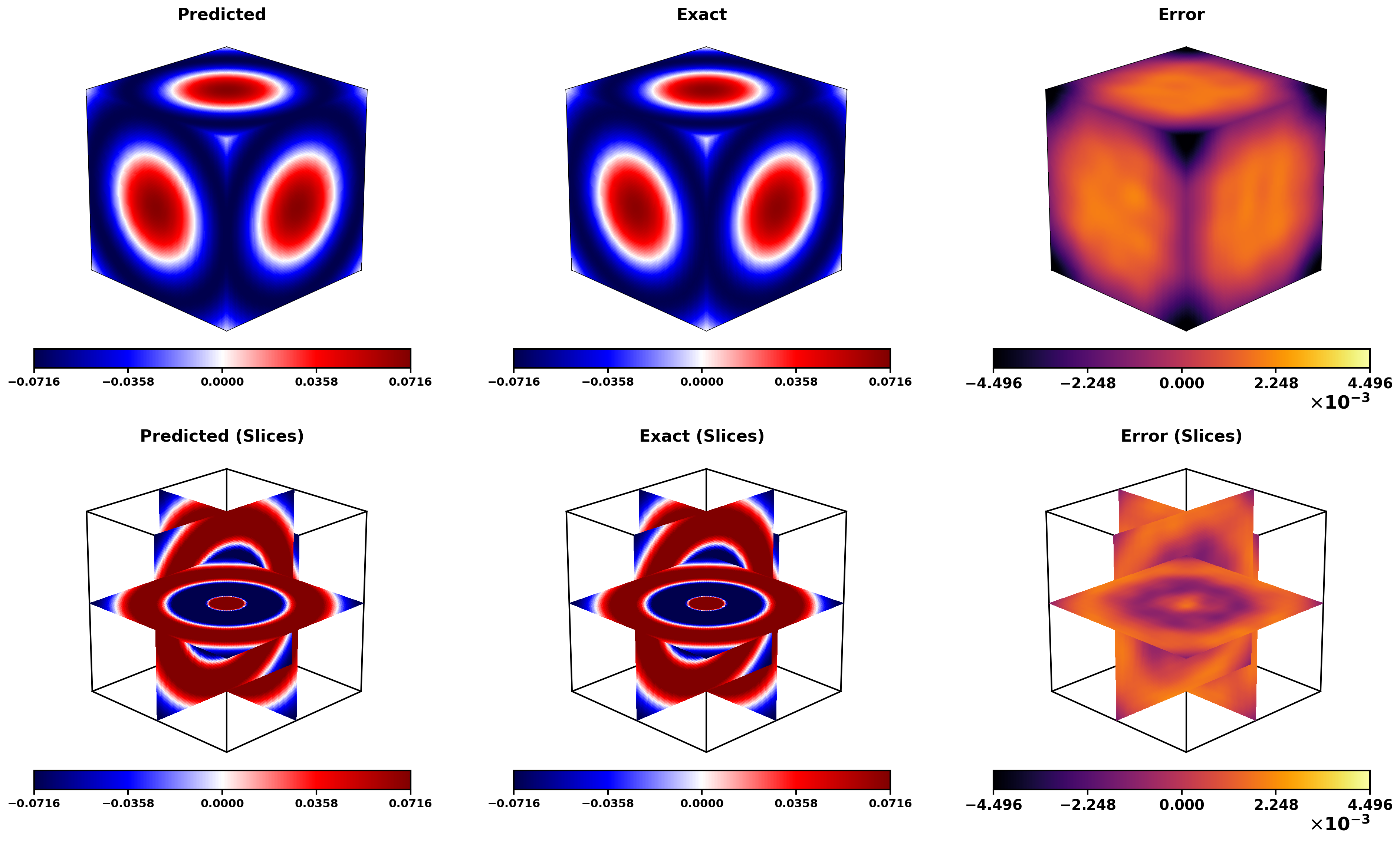}
   \caption{3D  radial velocity gradient benchmark: real part of the  scattered field $u_s$  mapped onto the outer surfaces of a cube enclosing the domain (top row) and via internal cross-section slices (bottom row). Comparison shows the prediction (left), the exact reference solution (center), and the pointwise error (right). Relative $L^2$ error: $1.53\%$.}
    \label{fig:Exact_3D_predictions}
\end{figure}

We first validate the proposed NAP-Wave solver against the closed-form
scattered-field solutions of the 2D fluid-cylinder scattering and 3D linear
radial velocity gradient benchmarks. The results are displayed in
Figures~\ref{fig:Exact_2D_predictions}-\ref{fig:Exact_3D_predictions}. In the
2D case, the predicted field is visually indistinguishable from the reference
solution: the wavefronts, amplitude decay, and phase structure are reproduced
across the full domain, and the pointwise error remains several orders of
magnitude smaller than the field itself, concentrated mainly near the source
and scatterer, where the field gradients are largest, and decaying rapidly at
larger radii. The line profiles along fixed $x$, fixed $z$, and the diagonal
cut confirm this agreement, with predicted and reference curves overlapping
almost exactly, including in the low-amplitude oscillatory tails. The 2D
benchmark achieves a relative $L^2$ error of $0.97\%$ and GI residual loss of order $10^{-9}$ with the plane-wave basis function, confirming that the
network accurately resolves both the oscillatory structure and the smoothly
decaying envelope.

In the 3D case, the solver reproduces the alternating positive and negative
lobes of the reference field closely on the domain boundary $\partial\Omega$,
shown here on the six faces of the cubic domain. Unlike in the 2D case, the
pointwise error is not concentrated near the source but is instead spread more
broadly over the boundary, with local maxima near the cube edges and corners.
The 3D benchmark attains a relative $L^2$ error of $1.53\%$, showing the scalability of the NAP-Wave solver.

\subsubsection{Marmousi benchmark}

\begin{figure}[!h]
    \centering
    \includegraphics[width=1.0\linewidth]{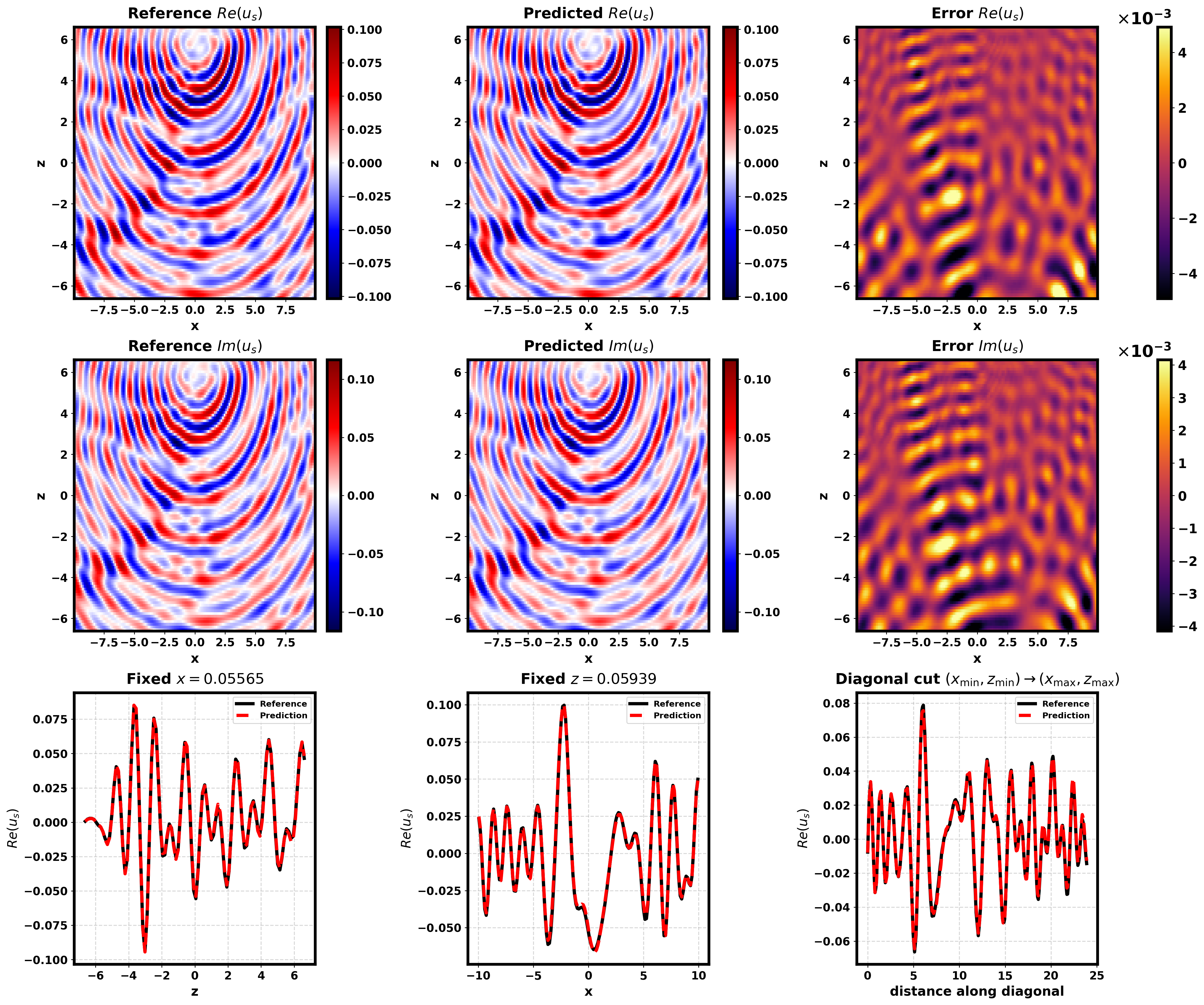}
    \caption{Marmousi benchmark: real part (top), imaginary part (middle) of $u_s$, and cross-sections (bottom).}
    \label{fig:Marmousi_predictions_full}
\end{figure}

The results obtained for the Marmousi benchmark are shown in
Figure~\ref{fig:Marmousi_predictions_full}. The predicted
scattered field reproduces the reference wavefronts across the domain, including their
pronounced curvature and the local wavefront distortions induced by the strong, finely
stratified velocity heterogeneity of the Marmousi model. The pointwise error of the real and
imaginary parts remains low over most of the domain, with the largest discrepancies concentrated
in the region of strongest lateral velocity contrast (roughly $x<-2.5$). The cross-sections
in Figure~\ref{fig:Marmousi_predictions_full} confirm close agreement between the predicted and
reference curves across the full oscillatory range.

\subsubsection{Overthrust benchmark}

\begin{figure}[!h]
    \centering
    \includegraphics[width=1.0\linewidth]{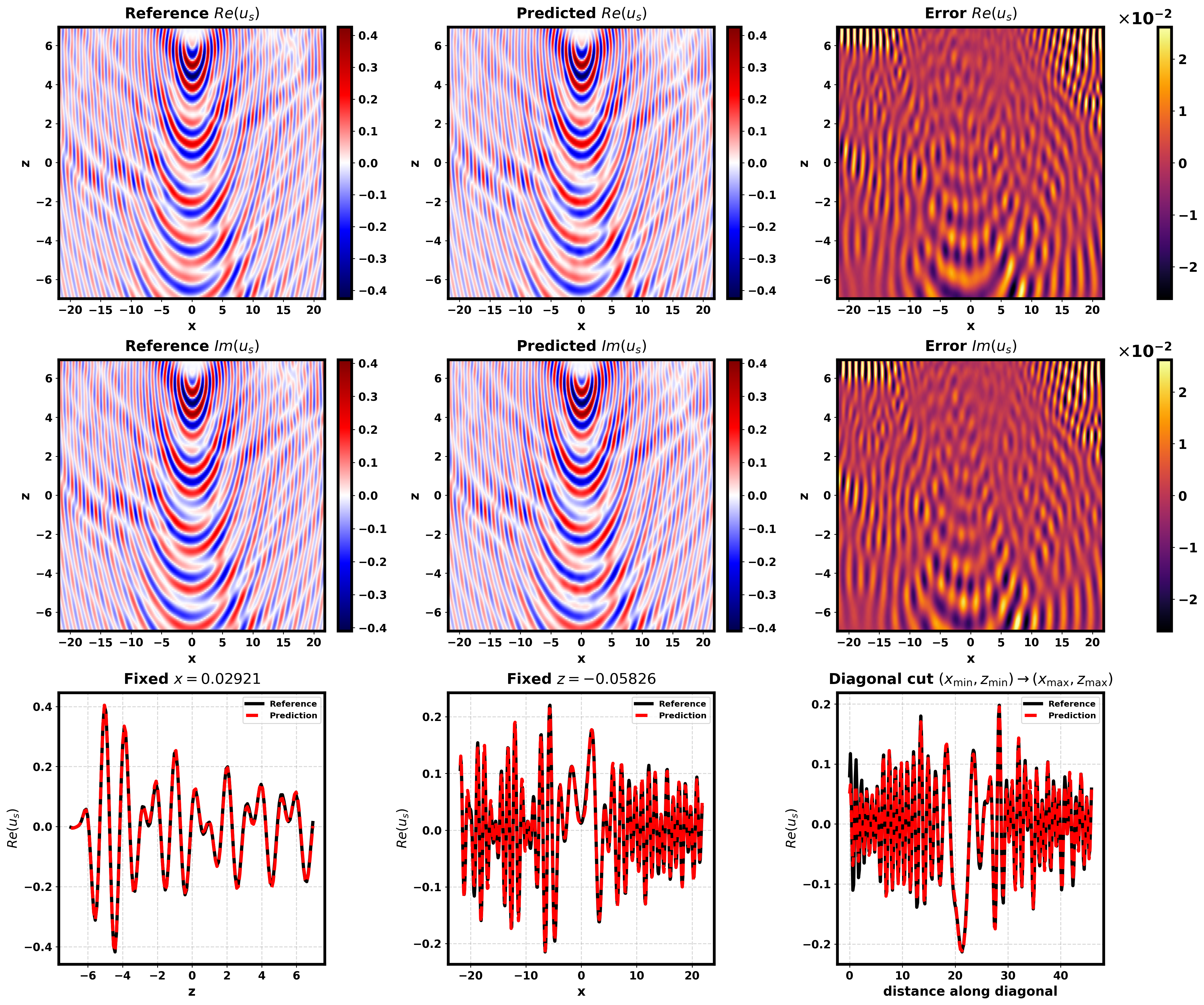}
    \caption{Overthrust benchmark: real part (top), imaginary part (middle) of $u_s$, and cross-sections (bottom).}
    \label{fig:Overthrust_predictions_full}
\end{figure}

We train the proposed NAP-Wave model on the Overthrust benchmark and  the results are
shown in Figure~\ref{fig:Overthrust_predictions_full}. The
predicted field correctly reproduces the sharp  wavefront pattern characteristic of
this benchmark, which arises from the abrupt fault-plane offsets in the Overthrust velocity model
acting as strong, spatially localized reflectors. Both the real and imaginary parts are in good quantitative agreement with the reference solution over the
bulk of the domain. The largest errors are concentrated near the top of the domain, close to the
source location and along the outer edge of the computational domain, where the truncated
finite-difference reference itself is most sensitive to residual reflections from the perfectly
matched layer. The cross-sections in Figure~\ref{fig:Overthrust_predictions_full} further confirm
that the model accurately resolves the multiscale character of the field, capturing both the
long-wavelength envelope and the fine oscillatory modes near the diagonal cut, although a modest
mismatch is visible at the very start of the diagonal profile.

\subsubsection{Otway benchmark}

\begin{figure}[!h]
    \centering
    \includegraphics[width=1.0\linewidth]{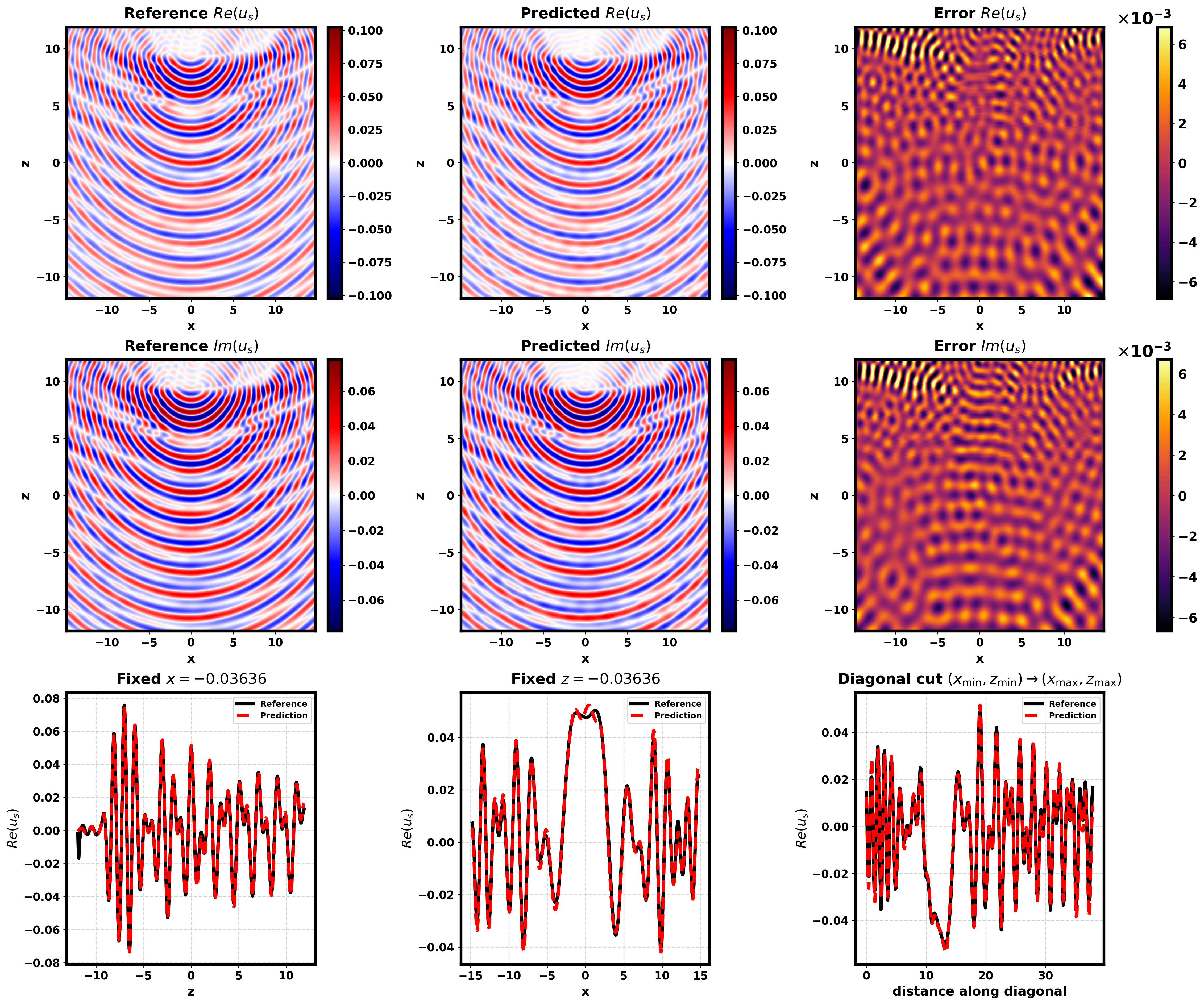}
    \caption{Otway benchmark: real part (top), imaginary part (middle) of $u_s$, and cross-sections (bottom).}
    \label{fig:Otway_predictions_full}
\end{figure}

The results obtained after training the NAP-Wave model on the Otway benchmark are
shown in Figure~\ref{fig:Otway_predictions_full}. Because the
Otway velocity model is comparatively smooth and gently layered relative to Marmousi and
Overthrust, the resulting scattered field exhibits smoother, more regular wavefronts, and the
network reproduces both the real and imaginary parts, as well as the amplitude, with pointwise
errors remaining close to zero over most of the domain. The residual error that does appear is
distributed as a fine, speckle-like pattern of low amplitude rather than being localized to a
specific region, consistent with a well-conditioned, mildly heterogeneous scattering problem. The cross-sections in
Figure~\ref{fig:Otway_predictions_full} confirm this good agreement, with the predicted and
reference curves nearly indistinguishable, including in the regions of most rapid oscillation.

\subsection{Comparison with the state of the art}
\label{subsec:sota}

We compare the performance of NAP-Wave against published results from the literature in Table~\ref{tab:comparison_literature}. The results show that NAP-Wave achieves the lowest  relative $L^2$ error on all three benchmarks compared to representative PINN methods.

\begin{table}[!h]
    \centering
    \caption{Comparison with representative PINN methods on Marmousi, Overthrust, and  Otway benchmarks.}
    \label{tab:comparison_literature}
    \vspace{4pt}
    \scriptsize
    \tablestyle
    \begin{tabular}{|p{2.4cm}|p{2.6cm}|p{1.6cm}|p{1.6cm}|p{1.6cm}|}
        \hline
        \rowcolor{tblheaderblue}
        \textbf{Reference} & \textbf{Method}
            & \textbf{Marmousi} \newline \textbf{Relative $L^2$}
            & \textbf{Overthrust} \newline \textbf{Relative $L^2$}
            & \textbf{Otway} \newline \textbf{Relative $L^2$} \\
        \hline
        Abedi et al.~\cite{abedi2025gabor} & Gabor PINN & 21.3\% & 21.1\% & not reported\\
        \hline
        \rowcolor{tblrowblue}
        Abedi et al.~\cite{abedi2026green} & Hybrid GI+PDE & 9.1\% & 13.2\% & 14.0\% \\
        \hline
        \rowcolor{tblrowblue}
        \textbf{NAP-Wave (ours)} & Neural Adaptive Plane-Wave  basis functions
       & $\mathbf{5}\%$ & $\mathbf{8.9}\%$  & $\mathbf{9.7}\%$ \\
        \hline
    \end{tabular}
\end{table}

\subsection{Ablation study}
\label{sec:ablation}

After we established that NAP-Wave produces accurate scattered fields across all
benchmarks, we now isolate the contribution of each design choice.

\subsubsection{Effect of the quadrature and basis size}
\label{subsec:ablation_exact}

\begin{figure}[!h]
    \centering
    \includegraphics[width=1.0\linewidth]{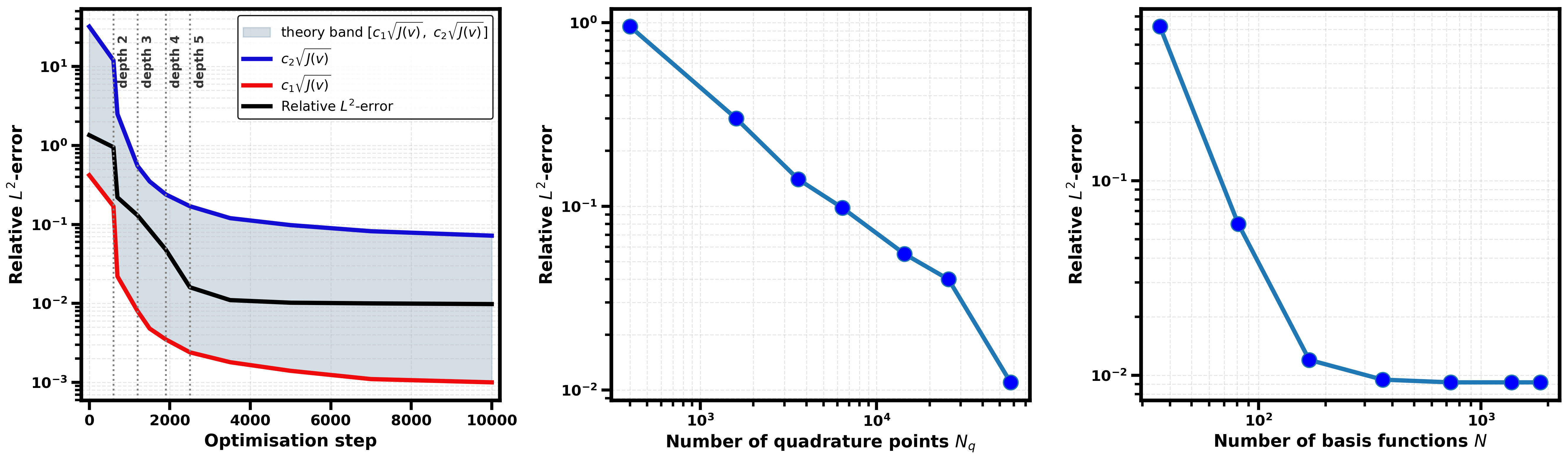}
    \caption{Fluid-cylinder scattering benchmark. Left: relative $L^2$ error versus optimization step with the theoretical bounds from Theorem~\ref{thm:error_estimate}; vertical dotted lines mark each depth transition. Middle: error versus $N_q$, with the number of basis functions fixed. Right: error versus $N$, with the number of quadrature points fixed.}
    \label{fig:Exact_solution2D_bound_check}
\end{figure}

We perform this ablation on the fluid-cylinder scattering benchmark. We study how the accuracy of the NAP-Wave solver depends on (i) the number of quadrature points used to approximate the Green's-integral loss, and (ii) the number of plane-wave neural basis functions. Accuracy is measured by the relative $L^2$ error against the exact solution and by the Green's-integral loss during training. We also check whether the theoretical bound relating the
Green's-integral loss to the true error, given in Theorem~\ref{thm:error_estimate}, holds in
practice at the discrete level as training progresses. Figure~\ref{fig:Exact_solution2D_bound_check}
summarizes these three results.

\paragraph{Error bound versus training progress.}
The left panel in Figure~\ref{fig:Exact_solution2D_bound_check} tracks the relative $L^2$ error (black) over the course of training,
together with the theoretical lower and upper bounds $c_1\sqrt{J(u_\theta)}$ and
$c_2\sqrt{J(u_\theta)}$ derived from the Green's-integral loss $J(u_\theta)$. The shaded region is the range
allowed by the theory, and the vertical dotted lines mark the steps at which a new layer is added
under the adaptive-depth training strategy. The true error stays inside the theoretical band
throughout training, including across every depth transition, and both quantities flatten out
together once the network has converged at depth 5, confirming the bound in this setting.

\paragraph{Effect of the number of quadrature points.}
The middle panel in Figure~\ref{fig:Exact_solution2D_bound_check} shows the relative $L^2$ error as the number of Green's-integral evaluation
points $N_q$ is increased, with the number of basis functions held fixed. The error
decreases steadily, close to log-linearly, as the integral is evaluated more accurately, with no
sign of saturation over the range tested.

\paragraph{Effect of the number of basis functions.}
The right panel in Figure~\ref{fig:Exact_solution2D_bound_check} shows the relative $L^2$ error as the number of basis functions increases, with the number of quadrature points held fixed. The basis count is increased by growing the depth of the network. We observe that the relative $L^2$   error drops sharply at first and then plateaus beyond roughly $200$ basis functions. One explanation is that, beyond a certain network size, the error becomes limited by other factors, such as the discretization of the integral, rather than by the number of basis functions. Another possible explanation is that as more basis functions are added, they become increasingly linearly dependent and start to carry redundant information, so that further increases no longer improve the approximation.

\subsubsection{Effect of the depth schedule, the coefficient solve, and the basis choice}
\label{subsec:ablation_marmousi}
We consider the Marmousi model to test the robustness of the proposed NAP-Wave solver through three
ablations: the depth-training schedule, the linear-coefficient optimization strategy,
and the choice of basis function.

\begin{figure}[!h]
\centering
\includegraphics[width=\linewidth]{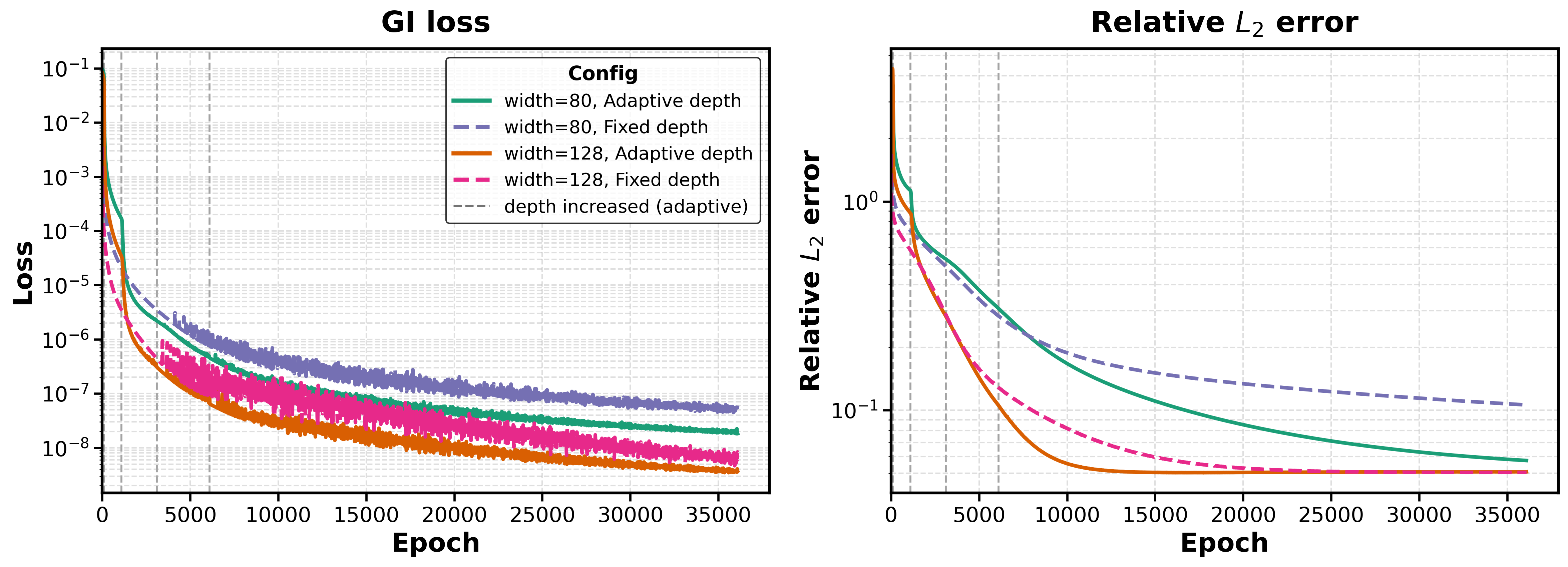}
      \label{fig:plane_wave_depth_loss}
\caption{Marmousi benchmark: effect of the depth-training schedule (adaptive vs.\ fixed) and
network width on convergence, for the plane-wave basis, with
$\bm{c}$ solved by least squares throughout. The GI loss (left) and relative $L^2$ error (right) are plotted against
optimization step on a
log scale. Vertical dashed lines mark the steps at which a new layer is added.}
\label{fig:marmousi_depth_ablation}
\end{figure}

\paragraph{Depth-training schedule.}
We compare the adaptive-depth training strategy against training with a fixed depth from the
start, for final depths ranging from $2$ to $6$, using the plane-wave basis with $\bm{c}$ solved
by least squares, under an equal total iteration budget $T_{\mathrm{total}}(L) = 1000 \times L$
for fairness, and averaging each configuration over $n=5$ independent seeds. Results are reported
in Table~\ref{tab:plane_wave_fixed_adaptive_depth_ablation} and
Figure~\ref{fig:marmousi_depth_ablation}. Table~\ref{tab:plane_wave_fixed_adaptive_depth_ablation}
shows that adaptive-depth is more accurate than fixed depth at every depth in
relative $L^2$ error, and more accurate in GI loss as well from depth $3$ onward. Moreover, adaptive depth is also faster in total training time at every depth,
by an average of about $1.21\times$ across depths $2$-$6$. Figure~\ref{fig:marmousi_depth_ablation}
extends this comparison to widths $80$ and $128$, up to depth $5$. The ranking is consistent across all four settings: at both widths, the
adaptive schedule (solid) reaches a given loss or error level at fewer
optimization steps than the fixed schedule (dashed), and the gap opens
immediately after the first layer is added. The narrower network (80, green)
plateaus at a visibly higher GI loss and relative $L^2$ error than the wider
one (128, orange) under either schedule.

\begin{table}[!h]
    \centering
    \small
\caption{Marmousi benchmark: effect of network depth and the adaptive-depth
schedule. Adaptive models follow the per-stage schedule $(T_1,\dots,T_L)$, and 
fixed models train in one stage for $T_{\mathrm{total}}(L)=\sum_\ell T_\ell$. Mean $\pm$ standard deviation over $n=5$ seeds. Bold
marks the better of Adaptive/Fixed per metric.}
\label{tab:plane_wave_fixed_adaptive_depth_ablation}
    \vspace{4pt}
    \tablestyle
    \small
    \begin{tabular}{|c|ccc|ccc|}
        \hline
        \rowcolor{tblheaderblue}
        & \multicolumn{3}{c|}{\textbf{Adaptive}} & \multicolumn{3}{c|}{\textbf{Fixed}} \\
        \rowcolor{tblheaderblue}
        \textbf{Depth} & \textbf{GI Loss} & \textbf{Rel.\ $L^2$ (\%)} & \textbf{Time (min)}
                        & \textbf{GI Loss} & \textbf{Rel.\ $L^2$ (\%)} & \textbf{Time (min)} \\
        \hline
        2 & $(1.186 \pm 0.156)\!\times\!10^{-5}$ & \textbf{76.02 $\pm$ 2.82} & \textbf{0.31 $\pm$ 0.00}
          & $\mathbf{(1.145 \pm 0.057)\!\times\!10^{-5}}$ & 79.61 $\pm$ 3.28 & 0.34 $\pm$ 0.01 \\
        \hline
        \rowcolor{tblrowblue}
        3 & $\mathbf{(3.082 \pm 0.098)\!\times\!10^{-7}}$ & \textbf{27.01 $\pm$ 1.29} & \textbf{0.62 $\pm$ 0.00}
          & $(1.361 \pm 0.085)\!\times\!10^{-6}$ & 43.26 $\pm$ 2.45 & 0.73 $\pm$ 0.01 \\
        \hline
        4 & $\mathbf{(1.233 \pm 0.083)\!\times\!10^{-7}}$ & \textbf{16.29 $\pm$ 0.83} & \textbf{1.05 $\pm$ 0.00}
          & $(3.375 \pm 0.595)\!\times\!10^{-7}$ & 21.20 $\pm$ 0.93 & 1.28 $\pm$ 0.01 \\
        \hline
        \rowcolor{tblrowblue}
        5 & $\mathbf{(7.502 \pm 2.858)\!\times\!10^{-8}}$ & \textbf{11.44 $\pm$ 0.64} & \textbf{1.59 $\pm$ 0.00}
          & $(1.575 \pm 0.542)\!\times\!10^{-7}$ & 14.42 $\pm$ 0.88 & 2.01 $\pm$ 0.01 \\
        \hline
        6 & $\mathbf{(3.006 \pm 0.522)\!\times\!10^{-8}}$ & \textbf{9.61 $\pm$ 0.67} & \textbf{2.24 $\pm$ 0.01}
          & $(1.519 \pm 0.394)\!\times\!10^{-7}$ & 11.81 $\pm$ 0.21 & 2.90 $\pm$ 0.01 \\
        \hline
    \end{tabular}
    \end{table}

\begin{figure}[!h]
    \centering
    \includegraphics[width=1.0\linewidth]{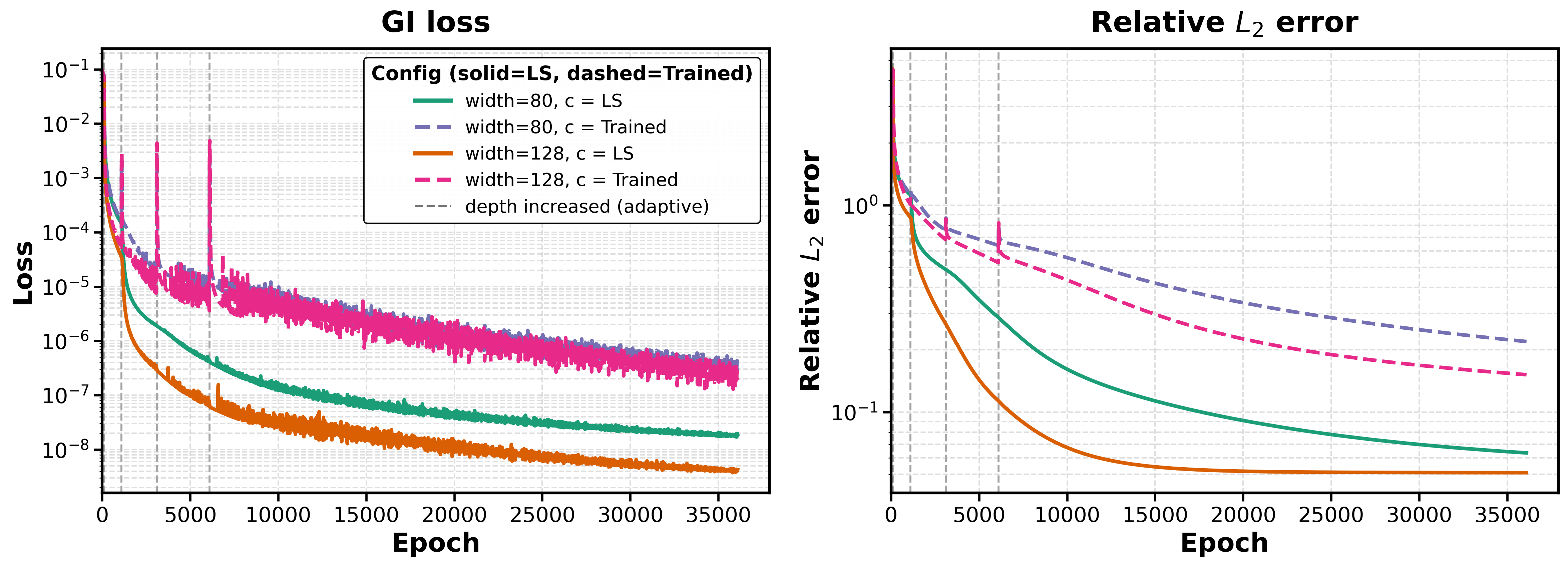}
    \caption{Marmousi benchmark: effect of the $\bm{c}$-update
    strategy on convergence, using the plane-wave basis and the adaptive-depth schedule. GI loss (left)
    and relative $L^2$ error (right) against optimization step, both on a log
    scale. Vertical dashed lines mark the steps at which a new
    layer is added.}
    \label{fig:marmousi_c_trajectory}
\end{figure}

\begin{figure}[!h]
\centering
\begin{subfigure}[t]{0.95\linewidth}
\centering
\includegraphics[width=\linewidth]{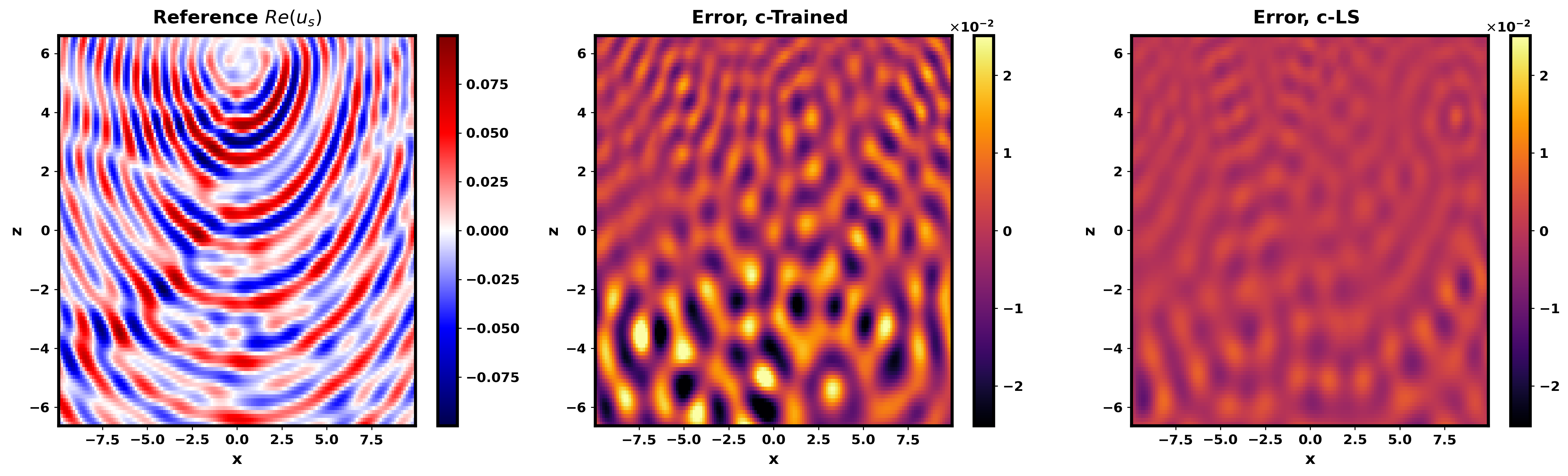}
      \label{fig:plane_wave_fields}
\end{subfigure}    \\[1em]
\caption{Marmousi benchmark. Effect of the $\bm{c}$-update strategy (closed-form least-squares
solve / trained jointly with $\bm{\theta}$) on the predicted scattered field $\mathrm{Re}(u_s)$,
using the adaptive-depth schedule with per-stage iteration table $(T_1,\dots,T_5) =
(100, 1000, 2000, 3000, 3000)$. Left: reference field. Middle: pointwise error for
\emph{c}-trained. Right: pointwise error for \emph{c}-LS, shared color scale with the
middle panel.}
\label{fig:marmousi_c_fields}
\end{figure}

\paragraph{Linear-coefficient optimization strategy.}
We now fix the depth-training schedule to adaptive and study how the strategy used for the linear
coefficients $\bm{c}$ affects convergence. Two strategies are compared: \emph{c}-LS, which solves
the normal equations \eqref{eq:discrete_normal_equations} exactly at each step for the current $\bm{\theta}$, and \emph{c}-trained,
which treats $\bm{c}$ as a trainable parameter updated jointly with $\bm{\theta}$ by Adam. In the \emph{c}-trained setting, coefficients introduced by a new layer are initialized
to zero so the augmented network initially reproduces the previous depth's output. Results are
reported in Figures~\ref{fig:marmousi_c_trajectory}-\ref{fig:marmousi_c_fields}.
Figure~\ref{fig:marmousi_c_trajectory} tracks the GI loss and the relative $L^2$ error against the optimization step for the plane-wave basis. The \emph{c}-LS variant attains a consistently lower GI loss and relative $L^2$ error than \emph{c}-trained throughout training, at both widths and at every depth transition. The GI-loss trajectory of \emph{c}-trained shows spikes at several depth transitions, most visibly at width $128$, before resuming its downward trend. The \emph{c}-LS trajectory shows no such spike at any of the depth transitions. We attribute this to two mechanisms. The coefficients $\bm{c}$ are solved exactly by least squares at each step, and each new layer is warm-started from the previous one, inheriting the optimizer's first- and second-moment estimates.  In Figure~\ref{fig:marmousi_c_fields}, we provide a qualitative visualization of this gap at the pointwise level. The \emph{c}-trained variant concentrates visibly larger pointwise error in the strongly heterogeneous region near $x < -5$, whereas \emph{c}-LS keeps a smaller and more uniformly distributed residual across the domain.

\begin{figure}[!h]
    \centering
\includegraphics[width=1.0\linewidth]{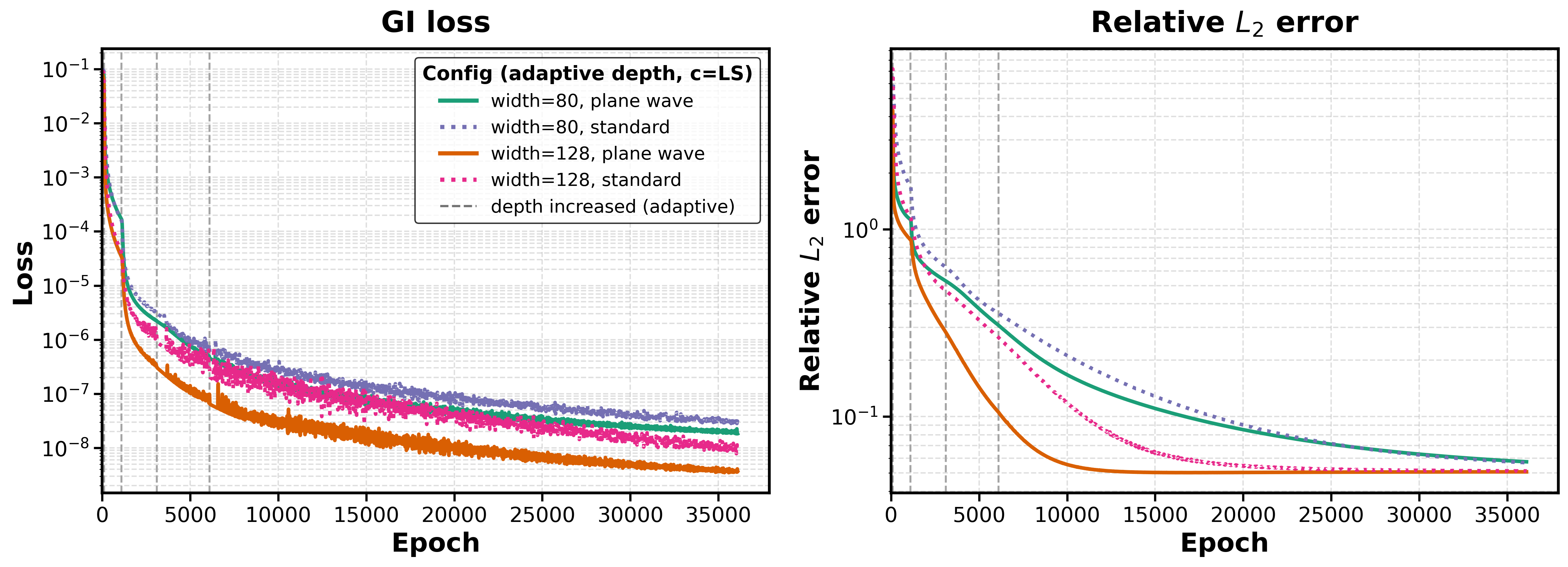}
    \caption{Marmousi benchmark, adaptive-depth schedule with $\bm{c}$ updated by least squares
    (\emph{c}-LS): effect of the plane-wave basis on convergence. GI loss (left) and relative
    $L^2$ error (right) against optimization step (Epoch), both on a log scale. Vertical dashed lines mark the steps at which a new
    layer is added.}
    \label{fig:plane_wave_vs_standard}
\end{figure}

\paragraph{Basis function choice.}
We retain the \emph{c}-LS strategy together with the adaptive-depth schedule, and compare the plane-wave and standard bases. Figure~\ref{fig:plane_wave_vs_standard} reports the results for widths $80$ and $128$. The plane-wave basis converges markedly faster: at width $128$ it reaches its final accuracy level after roughly $10{,}000$ epochs, whereas the standard basis needs the full budget of $35{,}000$ epochs to attain a comparable relative $L^2$ error. The two bases therefore reach a similar final $L^2$ accuracy, and the benefit of the plane-wave carrier is a large reduction in the number of iterations required. The GI loss behaves differently and remains consistently lower for the plane-wave basis throughout training, at both widths. We interpret this acceleration as a mitigation of spectral bias: the oscillatory component of the field is supplied by the carrier rather than learned, so the network only has to fit the slowly varying amplitude, whereas the standard basis must build the oscillations from scratch.

\paragraph{Combining all factors.}
We conclude with a quantitative ablation over all configurations, which combines the choice of basis, the coefficient strategy, and the depth schedule at two network widths. Table~\ref{tab:marmousi_threshold_crossing} reports, for each configuration, the optimization step and the wall-clock time at which the relative $L^2$ error first drops below a threshold of $10\%$. This measure isolates the cost of reaching an accurate solution rather than the accuracy attained at the end of a fixed budget. For every basis and every depth schedule, \emph{c}-LS crosses the threshold within about $20{,}000$ steps, whereas \emph{c}-trained requires at least $78{,}297$ steps, an acceleration of about four times. Under \emph{c}-LS, the plane-wave basis reaches the threshold before the standard basis at both widths and under both schedules, with a speedup of $1.5\times$ in wall-clock time. The adaptive-depth schedule adds a further speedup over fixed depth, which completes the picture: the three ingredients act in the same direction and their combination is the fastest configuration tested. Adaptive depth improves on fixed depth in almost every setting, \emph{c}-LS outperforms \emph{c}-trained in both GI loss and relative $L^2$ error regardless of basis, and the plane-wave basis reaches a given accuracy in fewer steps than a basis built from the network neurons alone. Together they reach $10\%$ error in $7{,}853$ steps and $2.30$~min.

\begin{table}[!h]
    \centering
    \small
\caption{Marmousi benchmark. Optimization step and wall-clock time at which each method's relative $L^2$ error first drops to or below a $10\%$ success threshold. All runs use depth $5$; the adaptive schedule spends 100, 1000, 2000, 3000 iterations at layers 1-4 respectively before growing, then trains the final (depth-5) layer until the threshold is hit. Every run is checked every epoch and stopped the instant its relative $L^2$ error crosses $10\%$. Bold marks the better of Adaptive/Fixed on each metric.}
\label{tab:marmousi_threshold_crossing}
    \vspace{4pt}
    \tablestyle
    \begin{tabular}{|c|c|cc|cc|}
        \hline
        \rowcolor{tblheaderblue}
        & & \multicolumn{2}{c|}{\textbf{Adaptive}} & \multicolumn{2}{c|}{\textbf{Fixed}} \\
        \rowcolor{tblheaderblue}
        \textbf{Basis / $c$-update} & \textbf{Width} & \textbf{Step to $10\%$} & \textbf{Time (min)}
                                     & \textbf{Step to $10\%$} & \textbf{Time (min)} \\
        \hline
        Plane-wave, $c$-LS & 80
          & \textbf{17,314} & \textbf{3.91} & 32,211 & 8.27 \\
        \hline
        \rowcolor{tblrowblue}
        Plane-wave, $c$-LS & 128
          & \textbf{7,853} & \textbf{2.30} & 8,228 & 3.32 \\
        \hline
        Plane-wave, $c$-trained & 80
          & \textbf{165,262} & \textbf{8.64} & 346,188 & 18.19 \\
        \hline
        \rowcolor{tblrowblue}
        Plane-wave, $c$-trained & 128
          & \textbf{105,917} & \textbf{7.67} & 119,746 & 8.81 \\
        \hline
        Standard, $c$-LS & 80
          & \textbf{20,033} & \textbf{4.47} & 39,637 & 9.89 \\
        \hline
        \rowcolor{tblrowblue}
        Standard, $c$-LS & 128
          & \textbf{10,918} & \textbf{3.45} & 11,952 & 4.71 \\
        \hline
        Standard, $c$-trained & 80
          & 149,557 & 6.74 & \textbf{143,182} & \textbf{6.55} \\
        \hline
        \rowcolor{tblrowblue}
        Standard, $c$-trained & 128
          & 88,308 & 5.34 & \textbf{78,297} & \textbf{4.85} \\
        \hline
    \end{tabular}
    \end{table}

\section{Conclusion}
\label{sec:conclusion}

We introduced NAP-Wave, a data-free physics-informed neural solver for time-harmonic scattering in heterogeneous media. The solver is trained by minimizing the residual of the Lippmann-Schwinger integral equation, which requires no differentiation in space of the network and makes training more stable and faster than strong-form PINN losses. Our main contributions are: (a) a neural adaptive plane-wave basis, where each basis function is the product of a neural amplitude and a plane-wave carrier with trainable wavenumber and fixed direction that solves the homogeneous Helmholtz equation exactly, encoding the oscillatory structure of the field directly into the trial space rather than learning it from scratch; (b) a decoupled optimization scheme, where the linear coefficients $\bm{c}$ are solved exactly at each step by least squares while the nonlinear basis parameters $\theta$ are updated by a gradient-based optimizer, yielding a better-conditioned optimization than training all parameters jointly; and (c) an adaptive-depth training strategy, where the network starts shallow and grows deeper as training progresses, which makes training stable and efficient.

We validated the proposed solver  against exact 2D and 3D solutions and on three realistic geophysical benchmarks (Marmousi 10~Hz, Overthrust 10~Hz, Otway 20~Hz), achieving relative $L^2$ errors of order $1\%$ on the 2D and 3D exact solutions, and $5\%$, $9\%$, and $10\%$ on Marmousi, Overthrust, and Otway respectively. An ablation study isolates the effect of the plane-wave basis and the exact least-squares coefficient solve: solving for $\bm{c}$ by least squares is markedly more accurate than training $\bm{c}$ and $\theta$ jointly, and the plane-wave basis leads to faster convergence. The adaptive-depth strategy makes training faster and more reliable than training at full depth from the start.

We note some limitations of this study and potential improvements. First, the Lippmann-Schwinger formulation assumes a constant background wave speed $v_0$, which enables the FFT-accelerated evaluation of the Green operator; a spatially varying background would break this structure and require a different formulation. Second, although the method was validated on a 3D exact solution, scaling to realistic 3D field-scale models proved difficult in practice due to the rapid growth of the memory footprint and the number of integration points and basis functions required; memory-efficient strategies, such as multi-GPU implementations, are needed to scale to full 3D models. Third, the method relies on a uniform quadrature grid and does not support local refinement; adaptive discretizations would address this, at the cost of losing the FFT structure. Fourth, the GI loss and the relative $L^2$ error do not decay at the same rate. The loss can reaches order $10^{-9}$ while the error stagnates around $10^{-2}$, a gap controlled by the conditioning of $\mathcal{B}$. An efficient preconditioner for $\mathcal{B}$, would align the two rates and may improve the final accuracy. Fifth, in the proposed adaptive-depth training strategy, the iteration budget allocated to each depth stage is fixed in advance rather than triggered adaptively from the loss; automating the depth schedule from a more faithful convergence signal is a promising direction. Sixth, NAP-Wave is a single-instance solver: any change of frequency or the velocity, model requires retraining from scratch; an extension toward operator learning would remove this restriction and, combined with the differentiability of the solver, would open the way to coupling it with inversion frameworks.

\section*{Acknowledgments}
The authors thank the DeepWave Consortium sponsors for supporting this research. 
This research was supported by the following Research Projects/Grants: 
European Union’s Horizon Europe research and innovation programme under the Marie Sklodowska-Curie Actions MSCA-DN-101119556 (IN-DEEP), and MSCA-SE-101342459 (MOST-PINN). 
PID2023-146678OB-I00 funded by MICIU/AEI /10.13039/501100011033 and by FEDER, EU;
BCAM Severo Ochoa accreditation of excellence CEX2021-001142-S funded by MICIU/AEI/10.13039/501100011033; 
Basque Government through the BERC 2022-2025 program;
EGOKIA (KK-2026/00050), funded by the Basque Government through ELKARTEK;
Consolidated Research Group MATHMODE (IT1866-26) of the UPV/EHU given by the Department of Sciences, Universities, and Innovation of the Basque Government; 
BCAM-IKUR-UPV/EHU, funded by the Basque Government IKUR Strategy and by the European Union NextGenerationEU/PRTR.

\section*{Code availability} 
The source code is publicly available at
\url{https://github.com/Hamza-Kamil/NAP-WAVE}.

\bibliographystyle{abbrvnat}  
\bibliography{references}

\begin{appendices}

\section{Notation}
\label{app:notation}

Table~\ref{tab:notation} collects the symbols used throughout the paper.

\begin{table}[!h]
    \centering
    \scriptsize
    \renewcommand{\arraystretch}{1.15}
\caption{Summary of notation used throughout the paper.}
\label{tab:notation}
    \vspace{4pt}
    \tablestyle
    \setlength{\tabcolsep}{3pt}
\begin{tabular}{|c|p{0.19\textwidth}|c|p{0.19\textwidth}|c|p{0.19\textwidth}|}
\hline
\rowcolor{tblheaderblue}
\multicolumn{2}{|c|}{\textbf{Problem setting}} &
\multicolumn{2}{c|}{\textbf{Network and trial space}} &
\multicolumn{2}{c|}{\textbf{Discretization and training}} \\
\hline
$d$ & Spatial dimension, $d=2,3$
  & $L$, $\ell$ & Target depth, current depth stage
  & $N_q$, $\{\bm{x}_k\}$ & Quadrature points, uniform grid \\
\hline
\rowcolor{tblrowblue}
$\Omega$, $\Omega_{\mathrm{full}}$, $\Gamma_w$ & Physical, padded domain, taper region
  & $\bm{a}=(a_0,\dots,a_L)$ & Architecture vector; $a_\ell$ width of layer $\ell$
  & $\Delta x$, $\Delta z$ & Grid spacings \\
\hline
$\omega$, $f_0$ & Angular, temporal frequency, $\omega=2\pi f_0$
  & $\theta\in\mathbb{R}^p$ & Network parameters; $p$ their number
  & $w_x$, $w_z$ & Taper widths, horizontal and vertical \\
\hline
\rowcolor{tblrowblue}
$v(\bm{x})$, $v_0$ & Velocity field, background velocity
  & $W^{(\ell)}$, $\bm{b}^{(\ell)}$, $\sigma$ & Weights, biases, activation of layer $\ell$
  & $J$, $J_\mu$, $J_\mu^{(N_q)}$ & Loss, reduced, empirical reduced loss \\
\hline
$k(\bm{x})$, $k_0$ & Local, background wavenumber, $k=\omega/v$
  & $N$, $N_\ell$ & Total basis, active basis at depth $\ell$
  & $\mu$, $\mu_n$ & Tikhonov parameter, decay schedule \\
\hline
\rowcolor{tblrowblue}
$m$, $m_0$, $\delta m$, $\widetilde{\delta m}$ & Squared slowness, background, contrast, tapered
  & $\varphi_j(\cdot\,;\theta)$ & $j$-th complex neural basis function
  & $T_\ell$, $T_{\mathrm{total}}$ & Iterations at stage $\ell$, total budget \\
\hline
$\lambda_0$ & Nondimensionalization length, $2\pi v_0/\omega$
  & $\bm{d}_j$, $k_j$, $\chi_j$ & Direction, wavenumber, carrier $e^{\mathrm{i}k_j\bm{d}_j\cdot\bm{x}}$
  & $\mathcal{F}$, $\mathcal{F}^{-1}$ & Discrete Fourier transform, inverse \\
\hline
\rowcolor{tblrowblue}
$\bm{x}_s$, $u_0$, $u_s$ & Source position, incident, scattered field
  & $k_{\min}$ & Lower bound of trainable wavenumber
  & $\odot$, $*$, $\circledast$ & Pointwise, linear, circular convolution \\
\hline
$g_0$, $G_0$ & Free-space Green's kernel and function
  & $\mathcal{V}_N(\theta)$, $\mathcal{V}$, $\mathcal{V}_w$ & Trial space, standard, plane-wave classes
  & & \\
\hline
\rowcolor{tblrowblue}
$\mathcal{G}$, $\mathcal{B}$ & Green integral operator, $\mathcal{B}=I-\mathcal{G}$
  & $\bm{c}$, $\bm{c}_\mu(\theta)$ & Linear coefficients, regularized solution
  & & \\
\hline
$f$ & Right-hand side, $f=\mathcal{G}u_0$
  & $\Phi_\theta$, $\mathsf{B}_\theta$, $\mathsf{F}_\theta$ & Basis, operator, weighted-basis matrices
  & & \\
\hline
\end{tabular}
\end{table}

\section{Neural network approximation}
\label{appendix:arch}

We present the NN architecture used to approximate the unknown
scattered field $u_s \in L^2(\Omega)$, together with the approximation spaces
it generates.

\subsection{Forward pass}
\label{sec:forward_pass}

Let $L \in \mathbb{N}$ denote the network depth and let $\bm{a} = (a_0, a_1,
\dots, a_L)$ be the architecture vector, where $a_0$ is the input dimension and
$a_\ell \in 2\mathbb{N}$ is the width of the $\ell$-th hidden layer. The
parameter space associated with $\bm{a}$ is
\begin{equation}
\label{eq:parameter_space}
    \Theta_{\bm{a}}
    =
    \prod_{\ell=1}^L
    \bigl(\RR^{a_\ell \times a_{\ell-1}} \times \RR^{a_\ell}\bigr),
\end{equation}
which is identified with $\RR^p$ where $p = \sum_{\ell=1}^L a_\ell(a_{\ell-1}+1)$
is the total number of trainable parameters. Fixing the architecture $\bm{a}$
also fixes the number of basis functions, given by $N = \sum_{\ell=1}^L
a_\ell/2$, consistently with the notation of Section~\ref{sec:approximation_space}.

For fixed $\theta = \{W_\ell, b_\ell\}_{\ell=1}^L \in \Theta_{\bm{a}}$, the
NN realization $u(\,\cdot\,;\theta) : \Omega \to
\RR^{a_L}$ is defined by the composition
\begin{equation}
\label{eq:network_realization}
   u(x;\theta)
    =
    h_L \circ \cdots \circ h_1 (x),
    \qquad x \in \Omega,
\end{equation}
where each layer map
$h_\ell : \RR^{a_{\ell-1}} \to \RR^{a_\ell}$ is given by
\begin{equation}
\label{eq:layer_map}
    h_\ell(\bm{z}) = \sigma(W_\ell \bm{z} + b_\ell),
    \qquad \bm{z} \in \RR^{a_{\ell-1}},
\end{equation}
with $W_\ell \in \RR^{a_\ell \times a_{\ell-1}}$, $b_\ell \in
\RR^{a_\ell}$, and $\sigma : \RR \to \RR$ an activation
function applied componentwise.

\subsection{Neural approximation space}
\label{sec:neural_space}

Since $u_s$ is complex-valued, we extract complex basis functions from the
real-valued hidden features of $u(\,\cdot\,;\theta)$ by pairing
consecutive neurons. For each layer $\ell = 1, \dots, L$ and each index $j =
1, \dots, a_\ell/2$, two constructions are considered.

\textbf{Standard basis.} The basis functions are defined directly from the
hidden features:
\begin{equation}
\label{eq:complex_basis_all_layers}
    \varphi_{\ell,j}(x;\theta)
    =
    h^{(\ell)}_{2j-1}(x;\theta)
    +
    \mathrm{i}\,h^{(\ell)}_{2j}(x;\theta).
\end{equation}

\textbf{Plane-wave basis.} Given a unit direction $\bm{d}_j \in \RR^d$ and
wavenumber $k_j > 0$, the basis functions are
\begin{equation}
\label{eq:plane_wave_basis}
    \varphi_{\ell,j}(x;\theta)
    =
    \bigl(h^{(\ell)}_{2j-1}(x;\theta)
    +
    \mathrm{i}\,h^{(\ell)}_{2j}(x;\theta)\bigr)
    \exp\!\left(\mathrm{i}\;k_j\;\bm{d}_j \cdot x\right).
\end{equation}
For fixed $\theta \in \Theta_{\bm{a}}$, the finite-dimensional neural space is
\begin{equation}
\label{eq:Vd_space}
    V_{\bm{a}}(\theta)
    =
    \operatorname{span}
    \left\{
        \varphi_{\ell,j}(\,\cdot\,;\theta)
        :
        \ell = 1, \dots, L,\;
        j = 1, \dots, \tfrac{a_\ell}{2}
    \right\},
\end{equation}
which has dimension at most $\sum_{\ell=1}^L a_\ell/2$. As $\theta$ ranges
over $\Theta_{\bm{a}}$, the union
\begin{equation}
\label{eq:Vd_union}
    \mathcal{V}_{\bm{a}}
    =
    \bigcup_{\theta \in \Theta_{\bm{a}}} V_{\bm{a}}(\theta)
\end{equation}
forms the search space for a fixed architecture. The full approximation class
over all admissible architectures is
\begin{equation}
\label{eq:V_full_space}
    \mathcal{V}
    =
    \bigcup_{\bm{a} \in \mathcal{A}} \mathcal{V}_{\bm{a}},
\end{equation}
where $\mathcal{A} \subset \bigcup_{L \in \mathbb{N}} \mathbb{N}^L$ denotes
the set of all architecture vectors with even widths. 

\section{Proofs of theoretical results}
\label{sec:theoretical_results}

\subsection{Proof of Proposition~\ref{prop:invertibility_B}}
\label{proof:invertibility_B}

To establish a proof of Proposition~\ref{prop:invertibility_B}, we first need to show the following lemma.

\begin{lemma}[Hilbert-Schmidt property of $\mathcal{G}$]
\label{lemma:K_is_Hilbert_Schmidt}
Let $\Omega \subset \mathbb{R}^d$, $d \in \{2,3\}$, be a bounded domain with
$\delta m \in L^\infty(\Omega)$. Then $\mathcal{G} : L^2(\Omega) \to L^2(\Omega)$
is a Hilbert-Schmidt operator with kernel $K(x,y) = \omega^2 g_0(x-y)\delta m(y)$.
In particular, $\mathcal{G}$ is compact and bounded on $L^2(\Omega)$.
\end{lemma}

\begin{proof}
It suffices to show that $K \in L^2(\Omega \times \Omega)$. Using the definition of the kernel
and the boundedness of the contrast function,
\begin{equation}
\label{eq:K_majoration_1}
    \iint_{\Omega \times \Omega} |K(x,y)|^2 \,\mathrm{d}x\,\mathrm{d}y
    \leq \omega^4 \|\delta m\|_{L^\infty(\Omega)}^2
    \int_\Omega \left( \int_\Omega |g_0(x-y)|^2 \,\mathrm{d}x \right) \,\mathrm{d}y.
\end{equation}
For any fixed $y \in \Omega$, the domain $\Omega$ is contained in a closed ball $B_R(y)$ of radius
$R := \sup_{x,x' \in \Omega} \|x - x'\| < \infty$, so that
\begin{equation}
\label{eq:inner_bound}
    \int_\Omega |g_0(x-y)|^2 \,\mathrm{d}x \leq \int_{B_R(y)} |g_0(x-y)|^2 \,\mathrm{d}x
    = \int_{B_R(0)} |g_0(z)|^2 \,\mathrm{d}z,
\end{equation}
after the change of variables $z = x-y$. Passing to spherical coordinates
($r = \|z\|$, $\mathrm{d}z = r^{d-1}\,\mathrm{d}r\,\mathrm{d}S_{d-1}$, with $\mathrm{d}S_{d-1}$ the
surface measure on the unit sphere $S^{d-1}$), and writing $|S^{d-1}|$ for its total measure,
\begin{equation}
    \int_{B_R(0)} |g_0(z)|^2 \,\mathrm{d}z = |S^{d-1}| \int_0^R |g_0(r)|^2 \, r^{d-1}\,\mathrm{d}r,
\end{equation}
where, with a slight abuse of notation, $g_0(r) := g_0(x)$ for $\|x\| = r$, since $g_0$ depends only
on $\|x\|$ in both cases considered. We split this radial integral into a singular neighborhood of
the origin and a regular exterior part, for fixed $\varepsilon \in (0,R)$:
\begin{equation}
    \int_0^R |g_0(r)|^2 r^{d-1}\,\mathrm{d}r
    = \int_0^\varepsilon |g_0(r)|^2 r^{d-1}\,\mathrm{d}r
    + \int_\varepsilon^R |g_0(r)|^2 r^{d-1}\,\mathrm{d}r.
\end{equation}
The second integral is finite since $g_0$ is continuous (hence bounded) on the compact interval
$[\varepsilon, R]$. It remains to check that the first integral is finite, which depends on the
dimension through the singularity of $g_0$ at the origin.

\smallskip
\noindent\textbf{Case $d=2$.} By \eqref{eq:nD_free_space_kernel}, $g_0(r) = \frac{i}{4}H_0^{(2)}(k_0 r)$,
and the asymptotic behavior of the zero-order Hankel function of the second kind near the origin
satisfies $|H_0^{(2)}(k_0 r)|^2 \sim \frac{4}{\pi^2}\ln^2(k_0 r)$ as $r \to 0^+$. Hence,
\begin{equation}
    \int_0^\varepsilon |g_0(r)|^2 r \,\mathrm{d}r
    \sim \frac{1}{4\pi^2} \int_0^\varepsilon r \ln^2(k_0 r) \,\mathrm{d}r.
\end{equation}
Substituting $u = k_0 r$, this scales as $\frac{1}{4\pi^2 k_0^2}\int_0^{\varepsilon k_0} u \ln^2(u)\,\mathrm{d}u$.
Since $\lim_{u \to 0^+} u^2\ln^2(u) = 0$, the antiderivative
\begin{equation}
    \int_0^{\varepsilon k_0} u \ln^2(u)\,\mathrm{d}u
    = \left[\tfrac{1}{2}u^2\ln^2(u) - \tfrac{1}{2}u^2\ln(u) + \tfrac{1}{4}u^2\right]_0^{\varepsilon k_0}
    < \infty
\end{equation}
confirms the integral is bounded.

\smallskip
\noindent\textbf{Case $d=3$.} By \eqref{eq:nD_free_space_kernel},
$g_0(r) = \frac{e^{-ik_0 r}}{4\pi r}$, so $|g_0(r)|^2 = \frac{1}{16\pi^2 r^2}$, and
\begin{equation}
    \int_0^\varepsilon |g_0(r)|^2 r^2 \,\mathrm{d}r
    = \frac{1}{16\pi^2} \int_0^\varepsilon \mathrm{d}r
    = \frac{\varepsilon}{16\pi^2} < \infty.
\end{equation}
In this case the apparent singularity of $g_0$ at the origin is entirely compensated by the
volume factor $r^{d-1} = r^2$, so no logarithmic correction is needed.

\smallskip
In either case, $\int_0^R |g_0(r)|^2 r^{d-1}\,\mathrm{d}r =: C(k_0, R) < \infty$, and substituting
back into \eqref{eq:K_majoration_1} gives
\begin{equation}
    \|K\|_{L^2(\Omega \times \Omega)}^2
    \leq \omega^4 \|\delta m\|_{L^\infty(\Omega)}^2 \, |S^{d-1}| \, C(k_0,R) \, |\Omega| < \infty,
\end{equation}
where $|\Omega|$ denotes the Lebesgue measure of $\Omega$. Consequently $K \in L^2(\Omega \times \Omega)$,
so $\mathcal{G}$ is a Hilbert-Schmidt operator, and in particular compact and bounded on $L^2(\Omega)$.
\end{proof}

 We now establish the proof of Proposition \ref{prop:invertibility_B}.

By Lemma~\ref{lemma:K_is_Hilbert_Schmidt}, the operator
$\mathcal{G} : L^2(\Omega) \to L^2(\Omega)$ is Hilbert-Schmidt, hence compact.
Therefore $\mathcal{B} = \mathcal{I} - \mathcal{G}$ is a Fredholm operator of
index zero. It follows that $\mathcal{B}$ is invertible if
and only if it is injective. To show injectivity, suppose  $\mathcal{B}u = 0$ for some $u \in L^2(\Omega)$,
i.e., $u = \mathcal{G}u$. Define the volume potential
\begin{equation}
\label{eq:volume_potential_w}
    w(x)
    = \omega^2 \int_\Omega g_0(x - y)\,\delta m(y)\,u(y)
      \,\mathrm{d}y, \qquad x \in \RR^d.
\end{equation}
Set $\varphi = \omega^2\,\delta m \cdot u \in L^2(\Omega)$. By Theorem~7.11
in~\cite{kirsch2011introduction}, $w \in H^2_{\mathrm{loc}}(\RR^d)$, $w$
satisfies the Sommerfeld radiation condition, and $w$ is the
unique radiating solution of $\Delta w + k_0^2 w = -\varphi$ in $\RR^d$. Since
$\varphi = \omega^2\,\delta m \cdot u$ is supported in $\Omega$, we have
$\varphi \equiv 0$ in $\RR^d \setminus \overline{\Omega}$, and therefore
\begin{equation}
\label{eq:helmholtz_exterior}
    \Delta w + k_0^2 w = 0 \qquad \text{in }
    \RR^d \setminus \overline{\Omega}.
\end{equation}
That is, $w$ is a radiating solution of the homogeneous Helmholtz equation in the
exterior of $\overline{\Omega}$. The Sommerfeld radiation condition
implies that
$\displaystyle\lim_{R\to\infty}\int_{||x||=R}|w|^2\,\mathrm{d}s = 0$
(see Proof of Theorem~7.9 in~\cite{kirsch2011introduction}).
By Rellich's lemma (Theorem~7.5
in~\cite{kirsch2011introduction}),
\begin{equation}
\label{eq:w_zero_exterior}
    w(x) = 0 \qquad \text{for all } \|x\| \geq b,
\end{equation}
where $b > 0$ is chosen so that $\overline{\Omega} \subset B(0,b)$.

The restriction of $w$ to $\Omega$ satisfies $w|_\Omega = \mathcal{G}u = u$. Substituting $u = w|_\Omega$ into
the equation $\Delta w + k_0^2 w = -\omega^2\,\delta m \cdot u$ on $\Omega$ gives
$\Delta w + k_0^2 w + \omega^2\,\delta m \cdot w = 0$ on $\Omega$, which can be
written as $\Delta w + k_0^2\,n(x)\,w = 0$, where
$n(x) = 1 + \omega^2\,\delta m(x)/k_0^2$. Since $\delta m \equiv 0$
outside $\Omega$, we have $n(x) = 1$ for $x \in \RR^d \setminus \Omega$,
and therefore $w$ satisfies $\Delta w + k_0^2\,n(x)\,w = 0$ in all of $\RR^d$
with $n \in L^\infty(\RR^d)$. By~\eqref{eq:w_zero_exterior}, $w$ has compact
support, so $w \in H^2(\RR^d)$. The hypotheses of Theorem~7.7
in~\cite{kirsch2011introduction} are therefore satisfied, and the conclusion of
that theorem gives $w \equiv 0$ on $\RR^d$. In particular, $u = w|_\Omega = 0$,
which establishes the injectivity of $\mathcal{B}$.  \hfill\qedsymbol

\subsection{Proof of Theorem~\ref{thm:existence_uniqueness_minimizer}}
\label{proof:existence_uniqueness_minimizer}

We first expand $J$ into its quadratic form. For any $v \in L^2(\Omega)$,
\begin{equation}
\label{eq:J_quadratic}
    J(v)
    =
    \frac{1}{2}\langle \mathcal{B}^* \mathcal{B} v, v \rangle_{L^2(\Omega)}
    -
    \operatorname{Re}\langle v, \mathcal{B}^* f \rangle_{L^2(\Omega)}
    +
    \frac{1}{2}\|f\|_{L^2(\Omega)}^2,
\end{equation}
where $\mathcal{B}^*$ denotes the Hilbert-space adjoint of $\mathcal{B}$.
The Hessian of $J$ is therefore $\mathcal{B}^*\mathcal{B}$. Since $\mathcal{B}$
is invertible (Proposition~\ref{prop:invertibility_B}), it has trivial kernel, so
$\mathcal{B}^*\mathcal{B}$ is strictly positive definite:
\begin{equation}
\label{eq:coercivity}
    \langle \mathcal{B}^*\mathcal{B} v, v \rangle_{L^2(\Omega)}
    =
    \|\mathcal{B}v\|_{L^2(\Omega)}^2 > 0
    \qquad \forall\, v \neq 0.
\end{equation}
Hence $J$ is strictly convex.

It remains to identify the minimizer. By Proposition~\ref{prop:invertibility_B},
there exists a unique $u_s \in L^2(\Omega)$ satisfying $\mathcal{B}u_s = f$,
and evaluating~\eqref{eq:cost_function_general} at $u_s$ gives $J(u_s) = 0$.
Since $J(v) \geq 0$ for all $v \in L^2(\Omega)$, the element $u_s$ is a global
minimizer. Uniqueness follows from strict convexity: if $v \in L^2(\Omega)$
also satisfies $J(v) = 0$, then $\|\mathcal{B}v - f\|_{L^2(\Omega)} = 0$,
hence $\mathcal{B}v = f$, and the injectivity of $\mathcal{B}$ forces $v = u_s$.
Furthermore, because $B$ is a Fredholm operator, its range $\text{Im}(B)$ is a closed subspace of $L^2(\Omega)$. This topological closure guarantees that the residual cost function $J$ defined in \eqref{eq:cost_function_general} possesses a unique global minimum, which corresponds exactly to the unique solution of the Lippmann-Schwinger equation. \hfill\qedsymbol

\subsection{Proof of Theorem~\ref{thm:error_estimate}}
\label{proof:error_estimate}

Since $\mathcal B: L^2(\Omega) \to L^2(\Omega)$ is bounded, linear, and invertible, the open
mapping theorem guarantees that its inverse $\mathcal B^{-1}: L^2(\Omega) \to L^2(\Omega)$ is
also linear and bounded. We first observe that the residual and the error are related
through $\mathcal B$ by
\begin{equation}
\label{eq:residual_error_relation}
    r(v) = f - \mathcal Bv = \mathcal Bu - \mathcal Bv = \mathcal B(u - v) = \mathcal Be(v).
\end{equation}
\noindent\textit{Upper bound.} Applying $\mathcal B^{-1}$ to both sides of
\eqref{eq:residual_error_relation} and using the boundedness of $\mathcal B^{-1}$,
\begin{equation}
    \|e(v)\|_{L^2(\Omega)}
    = \|\mathcal B^{-1} r(v)\|_{L^2(\Omega)}
    \le \|\mathcal B^{-1}\|_{\mathcal{L}(L^2(\Omega))} \|r(v)\|_{L^2(\Omega)}.
\end{equation}
\noindent\textit{Lower bound.} From \eqref{eq:residual_error_relation} and the
boundedness of $B$,
\begin{equation}
    \|r(v)\|_{L^2(\Omega)}
    = \|\mathcal Be(v)\|_{L^2(\Omega)}
    \le \|\mathcal B\|_{\mathcal{L}(L^2(\Omega))} \|e(v)\|_{L^2(\Omega)}.
\end{equation}
Since $\mathcal B$ is invertible, $\|\mathcal B\|_{\mathcal{L}(L^2(\Omega))} > 0$; dividing yields
\begin{equation}
    \|\mathcal B\|_{\mathcal{L}(L^2(\Omega))}^{-1} \|r(v)\|_{L^2(\Omega)}
    \le \|e(v)\|_{L^2(\Omega)}.
\end{equation}
Combining the two inequalities, dividing by $\|u\|_{L^2(\Omega)}$, and using
$\|r(v)\|_{L^2(\Omega)} = \sqrt{2J(v)}$ gives \eqref{eq:loss_error_bound} with
\begin{equation}
    c_1 = \frac{\sqrt{2}}{\|\mathcal{B}\|_{\mathcal{L}(L^2(\Omega))}\,\|u\|_{L^2(\Omega)}},
    \qquad
    c_2 = \frac{\sqrt{2}\,\|\mathcal{B}^{-1}\|_{\mathcal{L}(L^2(\Omega))}}{\|u\|_{L^2(\Omega)}}.
\end{equation} \hfill\qedsymbol

\subsection{Proof of Theorem~\ref{th:universal_approximation}}
\label{proof:universal_approximation}

We define the space of one-hidden-layer networks as:
\begin{equation}
\mathcal{M}(\sigma) = \left\{ f(x) = \sum_{j=1}^N c_j \sigma(w_j \cdot x + b_j) : N \in \mathbb{N}, \, c_j \in \mathbb{C}, \, w_j \in \RR^d, \, b_j \in \RR \right\}.
\end{equation}

We start with the case of standard neural basis functions. Since $\mathcal{M}(\sigma)$ is dense in $L^p(\Omega)$ (refer, e.g., to \cite{petersen2024mathematical} Corollary 3.19), the density of $\mathcal{V}$ follows from the fact that $\mathcal{M}(\sigma) \subset \mathcal{V}$. Indeed, let $f \in \mathcal{M}(\sigma)$ be an arbitrary function with $N$ neurons, $f(x) = \sum_{j=1}^N c_j \sigma(w_j \cdot x + b_j)$. We define the architecture $d = (d_0, 2N)$. For each neuron $j \in \{1, \dots, N\}$, we configure the parameters $\theta \in \Theta_d$ such that the hidden layer features are identical for the pair $(2j-1, 2j)$:
\begin{equation}
h^{(1)}_{2j-1}(x;\theta) = \sigma(w_j \cdot x + b_j), \quad h^{(1)}_{2j}(x;\theta) = \sigma(w_j \cdot x + b_j).
\end{equation}

The basis function for this pair is then:
\begin{equation}
\varphi_{1,j}(x;\theta) = h^{(1)}_{2j-1}(x;\theta) + \mathrm{i} h^{(1)}_{2j}(x;\theta) = (1 + \mathrm{i}) \sigma(w_j \cdot x + b_j).
\end{equation}

Defining coefficients $\alpha_j = \dfrac{c_j}{1+\mathrm{i}}$, we get:
\begin{equation}
\sum_{j=1}^N \alpha_j \varphi_{1,j}(x;\theta) = \sum_{j=1}^N \alpha_j (1 + \mathrm{i}) \sigma(w_j \cdot x + b_j) = \sum_{j=1}^N c_j \sigma(w_j \cdot x + b_j) = f(x).
\end{equation}

Thus, $\mathcal{M}(\sigma) \subseteq V_d(\theta) \subset \mathcal{V}$. 

We now show that the space $\mathcal{V}_w$, generated by the plane-wave neural basis functions, is dense in $L^p(\Omega)$ for all $p \in [1, \infty)$. Let $f \in L^p(\Omega)$ and let $\varepsilon > 0$. Since $C(\overline{\Omega})$ is dense in $L^p(\Omega)$, there exists a function $g \in C(\overline{\Omega})$ such that:
$\| f - g \|_{L^p(\Omega)} < \frac{\varepsilon}{2}.$ For a given wavenumber $k > 0$ and direction vector $\mathbf{d} \in \mathbb{S}^{1}$, we define the function $\tilde{g}(x) = g(x) e^{-\mathrm{i} k \mathbf{d} \cdot x}\in C(\overline{\Omega})$. Because $\mathcal{M}(\sigma)$ is dense in $C(\overline{\Omega})$, there exists $\varphi \in \mathcal{V}$ such that:
$\| \tilde{g} - \varphi \|_{L^\infty(\Omega)} < \frac{\varepsilon}{2 |\Omega|^{1/p}},$
where $|\Omega|$ denotes the Lebesgue measure of the domain. Let $h(x) = \varphi(x) e^{\mathrm{i} k \mathbf{d} \cdot x}$. By definition, $h \in \mathcal{V}_w$. Notice that $|e^{\mathrm{i} k \mathbf{d} \cdot x}| = 1$. Consequently:
\begin{equation}
\| g - h \|_{L^p(\Omega)}^p = \int_\Omega |\tilde{g}(x) - \varphi(x)|^p \, dx \le |\Omega| \| \tilde{g} - \varphi \|_{L^\infty(\Omega)}^p < |\Omega| \left( \frac{\varepsilon}{2 |\Omega|^{1/p}} \right)^p = \left( \frac{\varepsilon}{2} \right)^p.
\end{equation}
Finally, applying the triangle inequality, we get:
\begin{equation}
\| f - h \|_{L^p(\Omega)} \le \| f - g \|_{L^p(\Omega)} + \| g - h \|_{L^p(\Omega)} < \varepsilon.
\end{equation}
Thus, $\mathcal{V}_w$ is dense in $L^p(\Omega)$.
\hfill\qedsymbol

\subsection{Proof of Proposition~\ref{prop:c_theta_continuous}}
\label{proof:c_theta_continuous}

Fix $\mu > 0$ and $\theta \in \RR^p$. The operator $(\mathcal{B}\Phi_\theta)^* \mathcal{B}\Phi_\theta
+ \mu I$ is positive definite and in particular injective. Since it acts on the
finite-dimensional space $\mathbb{C}^N$, it is invertible, and therefore the
map $\bm{c}_\mu(\theta) = \left((\mathcal{B}\Phi_\theta)^* \mathcal{B}\Phi_\theta + \mu I\right)^{-1}
(\mathcal{B}\Phi_\theta)^* f$ is well defined. Since $(\mathcal{B}\Phi_\theta)^* \mathcal{B}\Phi_\theta : \mathbb{C}^N \to \mathbb{C}^N$, it is identified
with a matrix in $\mathbb{C}^{N \times N}$, denoted again by $(\mathcal{B}\Phi_\theta)^*
\mathcal{B}\Phi_\theta$. By definition of $\mathcal{B}\Phi_\theta$, for any $\bm{c}
\in \mathbb{C}^N$ and $g \in L^2(\Omega)$,
\begin{equation}
    \langle \mathcal{B}\Phi_\theta \bm{c},\, g \rangle_{L^2(\Omega)}
    =
    \sum_{j=1}^N c_j\,
    \langle \mathcal{B}\varphi_j(\cdot\,;\theta),\, g \rangle_{L^2(\Omega)}.
\end{equation}
Since $\langle \mathcal{B}\Phi_\theta \bm{c}, g \rangle_{L^2(\Omega)} = \langle \bm{c},
(\mathcal{B}\Phi_\theta)^* g \rangle_{\mathbb{C}^N}$ for all $\bm{c} \in \mathbb{C}^N$, the
$j$-th component of $(\mathcal{B}\Phi_\theta)^* g$ is
\begin{equation}
    \left((\mathcal{B}\Phi_\theta)^* g\right)_j
    =
    \langle g,\, \mathcal{B}\varphi_j(\cdot\,;\theta) \rangle_{L^2(\Omega)},
    \qquad j = 1, \ldots, N.
\end{equation}
Taking $g = \mathcal{B}\Phi_\theta \bm{
e}_i$, where $\bm{e}_i$ is the $i$-th canonical
basis vector of $\mathbb{C}^N$, gives the $(i,j)$-entry of $(\mathcal{B}\Phi_\theta)^*
\mathcal{B}\Phi_\theta$:
\begin{equation}
\label{eq:AstarA_entries}
    \left((\mathcal{B}\Phi_\theta)^* \mathcal{B}\Phi_\theta\right)_{ij}
    =
    \langle \mathcal{B}\varphi_i(\cdot\,;\theta),\,
    \mathcal{B}\varphi_j(\cdot\,;\theta) \rangle_{L^2(\Omega)},
    \qquad i,j = 1, \ldots, N.
\end{equation}
Taking $g = f$ gives the $i$-th entry of $(\mathcal{B}\Phi_\theta)^* f$:
\begin{equation}
\label{eq:Astarf_entries}
    \left((\mathcal{B}\Phi_\theta)^* f\right)_i
    =
    \langle f,\, \mathcal{B}\varphi_i(\cdot\,;\theta) \rangle_{L^2(\Omega)},
    \qquad i = 1, \ldots, N.
\end{equation}
Since each $\varphi_j(\cdot\,;\theta)$ is continuously differentiable in
$\theta$ by assumption, and $\mathcal{B}$ is a bounded linear operator
independent of $\theta$, the maps $\theta \mapsto
\mathcal{B}\varphi_j(\cdot\,;\theta)$ are continuously differentiable.
Therefore, by~\eqref{eq:AstarA_entries} and~\eqref{eq:Astarf_entries}, the
entries of $(\mathcal{B}\Phi_\theta)^* \mathcal{B}\Phi_\theta$ and $(\mathcal{B}\Phi_\theta)^* f$ are continuously
differentiable functions of $\theta$, and the map $\theta \mapsto (\mathcal{B}\Phi_\theta)^*
\mathcal{B}\Phi_\theta + \mu I$ is continuously differentiable for all $\theta \in
\RR^p$. Moreover, the matrix inversion map $M \mapsto M^{-1}$ is
continuously differentiable on the open set of invertible matrices in
$\mathbb{C}^{N \times N}$, and since $(\mathcal{B}\Phi_\theta)^* \mathcal{B}\Phi_\theta + \mu I$ is
invertible for every $\theta \in \RR^p$, the map $\theta \mapsto
((\mathcal{B}\Phi_\theta)^* \mathcal{B}\Phi_\theta + \mu I)^{-1}$ is continuously differentiable by the
chain rule. Therefore $\bm{c}_\mu(\theta) = ((\mathcal{B}\Phi_\theta)^* \mathcal{B}\Phi_\theta + \mu
I)^{-1} (\mathcal{B}\Phi_\theta)^* f$ is continuously differentiable with respect to $\theta$,
as a product of continuously differentiable maps. \hfill\qedsymbol 

\subsection{Proof of Proposition~\ref{prop:gradient_reduced_loss}}
\label{proof:gradient_reduced_loss}

Define $g : \RR^p \to \RR^p \times \mathbb{C}^N$ by
$g(\theta) := (\theta, \bm{c}_\mu(\theta))$, then $\mathcal{J}_\mu =
J_\mu \circ g$. By Proposition~\ref{prop:c_theta_continuous}, $\bm{c}_\mu$ is
continuously differentiable, and therefore $g$ is continuously differentiable.
Since $J_\mu$ is also differentiable, the chain rule gives
\begin{equation}
\label{eq:chain_rule}
    \mathrm{Jac}_{\mathcal{J}_\mu}(\theta)
    =
    \mathrm{Jac}_{J_\mu}(g(\theta))\;
    \mathrm{Jac}_g(\theta).
\end{equation}
The Jacobian of $g$ is
\begin{equation}
    \mathrm{Jac}_g(\theta)
    =
    \begin{pmatrix}
        I_p \\
        \mathrm{Jac}_{\bm{c}_\mu}(\theta)
    \end{pmatrix}
    \in \RR^{(p+N) \times p},
\end{equation}
and the Jacobian of $J_\mu$ at $g(\theta)$ decomposes as
\begin{equation}
    \mathrm{Jac}_{J_\mu}(g(\theta))
    =
    \begin{pmatrix}
        \nabla_\theta J_\mu(\theta, \bm{c}_\mu(\theta))^\top
        &
        \nabla_{\bm{c}} J_\mu(\theta, \bm{c}_\mu(\theta))^\top
    \end{pmatrix}
    \in \RR^{1 \times (p+N)}.
\end{equation}
Substitution into~\eqref{eq:chain_rule} and multiplication of the two block
matrices yields
\begin{equation}
\label{eq:chain_rule_expanded}
    \nabla \mathcal{J}_\mu(\theta)
    =
    \nabla_\theta J_\mu(\theta, \bm{c}_\mu(\theta))
    +
    \mathrm{Jac}_{\bm{c}_\mu}(\theta)^\top
    \nabla_{\bm{c}} J_\mu(\theta, \bm{c}_\mu(\theta)).
\end{equation}
Since $\bm{c}_\mu(\theta)$ minimizes $\bm{c} \mapsto J_\mu(\theta, \bm{c})$,
the first-order optimality condition gives $\nabla_{\bm{c}} J_\mu(\theta,
\bm{c}_\mu(\theta)) = 0$, and~\eqref{eq:chain_rule_expanded} reduces to
$\nabla \mathcal{J}_\mu(\theta) = \nabla_\theta J_\mu(\theta,
\bm{c}_\mu(\theta))$. Since the regularization term
$\frac{\mu}{2}\|\bm{c}\|_{\mathbb{C}^N}^2$ does not depend on $\theta$,
$\nabla_\theta J_\mu = \nabla_\theta J$,
and~\eqref{eq:reduced_gradient} follows. \hfill\qedsymbol

\section{Regularization of the free-space Green's function singularity}
\label{appendix:Regularization of the free-space Green's function singularity}

The free-space Green's function $g_0$ for the  Helmholtz equation
has a logarithmic singularity at the origin. When the integral operator
$\mathcal{G}$ is discretized on a uniform grid, the value $g_0(0)$ must be
approximated by a suitable regularized quantity. An efficient strategy is to
replace the pointwise value $g_0(0)$ by the average of $g_0$ over a small domain
$D$ centered at the origin with area equal to the grid cell area
$\Delta x\,\Delta z$:
\begin{equation}
\label{eq:reg_average}
    g_0(0) \approx \frac{1}{|D|}\int_D g_0(x)\,\mathrm{d}A.
\end{equation}
We present two approaches.

\paragraph{Circular-cell regularization.}
The first choice is the disk $D = B(0,h)$, where $h$ is chosen so that
$\pi h^2 = \Delta x\,\Delta z$, that is,
\begin{equation}
    h = \sqrt{\frac{\Delta x\,\Delta z}{\pi}}.
\end{equation}
Substitution into~\eqref{eq:reg_average} and use of the local expansion of $g_0$
near the origin yield
\begin{equation}
    g_0(0) \approx \frac{1}{2\pi}\left(\log h + \log\frac{k_0}{2} + \gamma - \frac{1}{2}\right) + \frac{i}{4},
\end{equation}
where $\gamma$ denotes the Euler-Mascheroni constant.

\paragraph{Rectangular-cell regularization.}
The second choice is to take $D$ as the actual grid cell centered at the origin,
\begin{equation}
    D = \left[-\frac{\Delta x}{2}, \frac{\Delta x}{2}\right] \times \left[-\frac{\Delta z}{2}, \frac{\Delta z}{2}\right],
    \qquad |D| = \Delta x\,\Delta z.
\end{equation}
Substitution into~\eqref{eq:reg_average} and use of the local expansion of $g_0$
give
\begin{equation}
    g_0(0) \approx I + \frac{1}{2\pi}\left(\log\frac{k_0}{2} + \gamma\right) + \frac{i}{4},
\end{equation}
where $I$ is the logarithmic average over the rectangular cell,
\begin{equation}
    I = \frac{1}{\Delta x\,\Delta z}\int_D \frac{1}{2\pi}\log\|x\|\,\mathrm{d}A.
\end{equation}
This integral admits the closed-form expression
\begin{equation}
    I = \frac{1}{4\pi}\left(\log(\Delta x^2 + \Delta z^2) - 2\log 2 - 3 + \frac{\Delta x}{\Delta z}\arctan\frac{\Delta z}{\Delta x} + \frac{\Delta z}{\Delta x}\arctan\frac{\Delta x}{\Delta z}\right).
\end{equation}

\section{Computational complexity of Algorithm~\ref{alg:hybrid_optimization}}
\label{appendix:complexity}

We summarize the per-iteration cost of Algorithm~\ref{alg:hybrid_optimization}.
At each outer iteration the parameters $\theta$ are fixed, the discrete operator
$\mathsf{B}_\theta \in \mathbb{C}^{N_q \times N}$ is assembled, and the discrete
normal equations~\eqref{eq:discrete_normal_equations} are solved for the
coefficient vector before the Adam update of $\theta$. Throughout, $N_q$ denotes
the number of quadrature points, $N$ the number of basis functions, and $p$ the
number of network parameters. The assembly of $\mathsf{B}_\theta$ costs
\begin{equation}
    O\!\left(N_q p + N N_q^2\right)\ \text{(direct)},
    \qquad
    O\!\left(N_q p + N N_q \log N_q\right)\ \text{(FFT)},
\end{equation}
and the full iteration cost using FFT 
\begin{equation}
    O\!\left(N_q p + N N_q \log N_q + N^2 N_q + N^3\right).
\end{equation}

\begin{table}[!h]
    \centering
    \caption{Per-iteration computational cost and memory of the main operations
    in Algorithm~\ref{alg:hybrid_optimization}, using the FFT realization of the
    Green operator. Here $N_q$ is the number of quadrature points, $N$ the number
    of basis functions, and $p$ the number of network parameters.}
    \label{tab:complexity}
    \vspace{4pt}
    \tablestyle
    \begin{tabular}{|l|c|c|}
        \hline
        \rowcolor{tblheaderblue}
        \textbf{Operation} & \textbf{Cost} & \textbf{Memory} \\
        \hline
        Forward pass $\Phi_\theta$ & $O(N_q p)$ & $O(N_q N)$ \\
        \hline
        \rowcolor{tblrowblue}
        Green operator $\mathcal{G}\Phi_\theta$ (FFT) & $O(N N_q \log N_q)$ & $O(N_q N)$ \\
        \hline
        Assemble $\mathsf{B}_\theta = \Phi_\theta - \mathcal{G}\Phi_\theta$ & $O(N_q N)$ & $O(N_q N)$ \\
        \hline
        \rowcolor{tblrowblue}
        Normal matrix $\mathsf{B}_\theta^* \mathsf{B}_\theta$ & $O(N^2 N_q)$ & $O(N^2)$ \\
        \hline
        Right-hand side $\mathsf{B}_\theta^* \bm{f}$ & $O(N N_q)$ & $O(N)$ \\
        \hline
        \rowcolor{tblrowblue}
        Cholesky factorization & $O(N^3)$ & $O(N^2)$ \\
        \hline
        Triangular solves (two) & $O(N^2)$ & $O(N^2)$ \\
        \hline
        \rowcolor{tblrowblue}
        Adam update of $\theta$ & $O(p)$ & $O(p)$ \\
        \hline
    \end{tabular}
\end{table}

\section{Other implementation details}
\label{app:fft_implementation}

\subsection{Enforcing compact support of $\delta m$ via damping taper}
\label{appendix:compact_support}

The Lippmann-Schwinger formulation requires $\operatorname{supp}(\delta m)\subset
\Omega$. In practice,
the physical velocity model is defined on a finite grid and the
contrast $\delta m$ may be nonzero up to and including the boundary of the
computational domain. We enforce the compact support condition numerically by
applying a separable raised-cosine (Hann) taper that smoothly drives $\delta m$ to
zero in a prescribed width border region (refer to Figure~\ref{fig:tapering_delta_m}).

Let the computational grid have $n_x$ points in the horizontal direction and $n_z$
points in the vertical direction. We reserve a border of $w_x$ cells on each side
in $x$ and $w_z$ cells on each side in $z$, defining an interior region of size
$(n_z - 2w_z)\times(n_x - 2w_x)$. The taper is constructed as the outer product of
two one-dimensional windows,
\begin{equation}
    \tau_{k\ell} = \psi_z(k)\,\psi_x(\ell),
    \qquad k = 1, \ldots, n_z,
    \quad \ell = 1, \ldots, n_x,
\end{equation}
where each factor $\psi$ is a raised-cosine ramp: taking the $x$-direction as
representative,
\begin{equation}
    \psi_x(\ell) =
    \begin{cases}
        \frac{1}{2}\left(1 - \cos\frac{\pi(\ell-1)}{w_x}\right), & \ell = 1, \ldots, w_x, \\[2mm]
        1, & \ell = w_x+1, \ldots, n_x - w_x, \\[2mm]
        \frac{1}{2}\left(1 - \cos\frac{\pi(n_x-\ell)}{w_x}\right), & \ell = n_x - w_x + 1, \ldots, n_x,
    \end{cases}
\end{equation}
and $\psi_z(k)$ is defined analogously. The window equals one throughout the
interior and ramps smoothly to exactly zero at the grid boundary, so that
$\tau_{k\ell} = 0$ whenever $k \in \{1, n_z\}$ or $\ell \in \{1, n_x\}$. The
tapered contrast is then
\begin{equation}
    \widetilde{\delta m}(x_{k\ell}) = \tau_{k\ell}\,\delta m(x_{k\ell}),
\end{equation}
which satisfies $\widetilde{\delta m}\equiv 0$ on the border cells and hence
$\operatorname{supp}(\widetilde{\delta m})\subset\Omega_\text{full}$, as required. This
replaces the original contrast $\delta m$ in the weighted basis
matrix~\eqref{eq:weighted_basis} and therefore in the Green
operator~\eqref{eq:green_matrix}.

\subsection{FFT-accelerated evaluation}
\label{appendix:fft_acceleration}

The matrix product~\eqref{eq:green_matrix} consists of $N$ independent Toeplitz
matrix-vector products~\eqref{eq:green_column}. A direct evaluation of each product
costs $O(N_q^2)$, giving a total cost of $O(N N_q^2)$. this can be reduced to
$O(N N_q \log N_q)$ by embedding the Toeplitz product into a circulant one and
applying the convolution theorem, implementing the discrete Fourier transform and
its inverse using an FFT algorithm.

Since $g_0$ is Toeplitz with $2N_q - 1$ distinct entries
$t_{-(N_q-1)}, \ldots, t_{N_q-1}$, we set $M = 2N_q - 1$ and pad both the kernel
and each column of $F_\theta$ to length $M$. Define the padded kernel
$\tilde{g}\in\mathbb{C}^M$ by
\begin{equation}
    \tilde{g}_r =
    \begin{cases}
        t_{r-1}, & r = 1, \ldots, N_q, \\
        t_{r-1-M}, & r = N_q + 1, \ldots, M,
    \end{cases}
\end{equation}
which packs all $2N_q - 1$ values into a single period of length $M$: the
non-negative lags $t_0, \ldots, t_{N_q-1}$ occupy the first $N_q$ entries and the
negative lags $t_{-(N_q-1)}, \ldots, t_{-1}$ are placed at the end. This vector
depends only on the kernel $g_0$ and is the same for every column $j$. For each
$j = 1, \ldots, N$, define the padded column $\tilde{f}_j\in\mathbb{C}^M$ by
\begin{equation}
    (\tilde{f}_j)_r =
    \begin{cases}
        (F_\theta)_{r,j}, & r = 1, \ldots, N_q, \\
        0, & r = N_q + 1, \ldots, M.
    \end{cases}
\end{equation}
The circular convolution of length $M$ is
\begin{equation}
    (\tilde{g}\circledast\tilde{f}_j)_k = \sum_{r=1}^M \tilde{g}_{((k-r)\bmod M)+1}(\tilde{f}_j)_r,
    \qquad k = 1, \ldots, M,
\end{equation}
which defines a circulant matrix-vector product of size $M$. For
$k\in\{1,\ldots,N_q\}$, the zero-padding kills all terms with $r > N_q$, and the
construction of $\tilde{g}$ ensures that the circulant index resolves to
$t_{k-r}$ for every $r\in\{1,\ldots,N_q\}$, so
\begin{equation}
    (\tilde{g}\circledast\tilde{f}_j)_k = \sum_{\ell=1}^{N_q} t_{k-\ell}(F_\theta)_{\ell,j} = (\bm{g}_0\bm{f}_j)_k = (\mathcal{G}\Phi_\theta)_{k,j}.
\end{equation}
That is, the first $N_q$ entries of the length-$M$ circular convolution reproduce
the Toeplitz product~\eqref{eq:green_column} exactly, with no wraparound. By the
convolution theorem
\begin{equation}
    \tilde{g}\circledast\tilde{f}_j = \mathcal{F}_M^{-1}\!\left(\mathcal{F}_M[\tilde{g}]\odot\mathcal{F}_M[\tilde{f}_j]\right),
\end{equation}
where $\mathcal{F}_M$ denotes the DFT of length $M$ and $\odot$ the pointwise
product. Combining the two, each column of $\mathcal{G}\Phi_\theta$ is recovered
by extracting the first $N_q$ entries,
\begin{equation}
    (\mathcal{G}\Phi_\theta)_{\cdot,j} = \left[\mathcal{F}_M^{-1}\!\left(\mathcal{F}_M[\tilde{g}]\odot\mathcal{F}_M[\tilde{f}_j]\right)\right]_{1:N_q},
    \qquad j = 1, \ldots, N,
\end{equation}
where $[\,\cdot\,]_{1:N_q}$ denotes extraction of the first $N_q$ entries. The
transform $\mathcal{F}_M[\tilde{g}]$ is computed once at cost $O(M\log M)$, and
each of the $N$ columns requires one forward FFT, one pointwise product, and one
inverse FFT, giving a total cost of $O(N M\log M) = O(N N_q\log N_q)$, compared
with $O(N N_q^2)$ for the direct Toeplitz products.

\end{appendices}

\end{document}